\documentstyle[psfig,12pt]{article}
\def\TL{\hfil$\displaystyle{##}$}
\def\TR{$\displaystyle{{}##}$\hfil}
\def\TC{\hfil$\displaystyle{##}$\hfil}
\def\TT{\hbox{##}}
\def\seqalign#1#2{\vcenter{\openup1\jot
  \halign{\strut #1\cr #2 \cr}}}

\def\fixit#1{}

\def\mop#1{\mathop{\rm #1}\nolimits}

\def\tr{\mop{tr}}

\def\overleftrightarrow#1{\vbox{\ialign{##\crcr
     $\leftrightarrow$\crcr\noalign{\kern-0pt\nointerlineskip}
     $\hfil\displaystyle{#1}\hfil$\crcr}}}

\def\lsim{\mathrel{\mathstrut\smash{\ooalign{\raise2.5pt\hbox{$<$}\cr\lower2.5pt\hbox{$\sim$}}}}}
\def\gsim{\mathrel{\mathstrut\smash{\ooalign{\raise2.5pt\hbox{$>$}\cr\lower2.5pt\hbox{$\sim$}}}}}

\def\slashed#1{{\ooalign{\hfil\hfil/\hfil\cr $#1$}}}

\def\sqr#1#2{{\vcenter{\vbox{\hrule height.#2pt
         \hbox{\vrule width.#2pt height#1pt \kern#1pt
            \vrule width.#2pt}
         \hrule height.#2pt}}}}

\def\href#1#2{#2}  

\def\lbldef#1#2{\expandafter\gdef\csname #1\endcsname {#2}}
\def\eqn#1#2{\lbldef{#1}{(\ref{#1})}%
\begin{equation} #2 \label{#1} \end{equation}}
\def\eqalign#1{\vcenter{\openup1\jot
    \halign{\strut\span\TL & \span\TR\cr #1 \cr
   }}}

\def\Re{\mop{Re}}
\def\Im{\mop{Im}}
\begin{document}
\baselineskip=15.5pt
\pagestyle{plain}
\setcounter{page}{1}

\begin{titlepage}

\begin{flushright}
CALT-68-2370 \\
CITUSC/02-001 \\
hep-th/0201114
\end{flushright}
\vfil

\begin{center}
{\huge TASI lectures: special holonomy}
\vskip0.3cm
{\huge in string theory and M-theory}
\end{center}

\vfil
\begin{center}
{\large Steven S. Gubser$^{\,}$\footnote{E-mail: 
{\tt ssgubser@theory.caltech.edu}}$^,$\footnote{On leave from Princeton
University.}}
\end{center}

$$\seqalign{\span\TL & \span\TT}{
& Lauritsen Laboratory of Physics, 452-48 Caltech, Pasadena, CA  91125
}$$
\vfil

\begin{center}
{\large Abstract}
\end{center}

\noindent 
 A brief, example-oriented introduction is given to special holonomy
and its uses in string theory and M-theory.  We discuss $A_k$
singularities and their resolution; the construction of a K3 surface
by resolving $T^4/{\bf Z}_2$; holomorphic cycles, calibrations, and
worldsheet instantons; aspects of the low-energy effective action for
string compactifications; the significance of the standard embedding
of the spin connection in the gauge group for heterotic string
compactifications; $G_2$ holonomy and its relation to ${\cal N}=1$
supersymmetric compactifications of M-theory; certain isolated $G_2$
singularities and their resolution; the Joyce construction of compact
manifolds of $G_2$ holonomy; the relation of D6-branes to M-theory on
special holonomy manifolds; gauge symmetry enhancement from light
wrapped M2-branes; and chiral fermions from intersecting branes.
These notes are based on lectures given at TASI~'01.

\vfil
\begin{flushleft}
January 2002
\end{flushleft}
\end{titlepage}
\newpage
\renewcommand{\thefootnote}{\arabic{footnote}}  
\setcounter{footnote}{0}
\section{Introduction}
\label{Introduction}

Special holonomy plays a prominent role in string theory and M-theory
primarily because the simplest vacua preserving some fraction of
supersymmetry are compactifications on manifolds of special holonomy.
The case that has received the most intensive study is Calabi-Yau
three-folds ($CY_3$), first because heterotic string compactifications
on such manifolds provided the first semi-realistic models of
particle phenomenology, and second because type II strings on
Calabi-Yau three-folds exhibit the seemingly miraculous property of
``mirror symmetry.''  Recently, seven-manifolds with $G_2$ holonomy
have received considerable attention, both because they provide the
simplest way to compactify M-theory to four dimensions with ${\cal
N}=1$ supersymmetry, and because of some unexpected connections with
strongly coupled gauge theory.

The purpose of these two lectures, delivered at TASI~'01, is to
introduce special holonomy in a way that will make minimal demands on
the reader's mathematical erudition,\footnote{It is my hope that a
graduate student who has learned General Relativity, knows the basic
facts about Lie groups and their representations, and has at least a
nodding acquaintance with string theory, will be able to follow the
gist of this presentation.  Some of the more advanced topics will
require more erudition or background reading.} but nevertheless get to
the point of appreciating a few deep facts about perturbative and
non-perturbative string theory.  Some disclaimers are in order: these
lectures do not aspire to mathematical rigor, nor to completeness.  I
have made a perhaps idiosyncratic selection of material that will
hopefully serve as a comprehensible invitation to the wider
literature.  To enhance the appeal of mathematical concepts that may
seem abstruse or dreary to the theoretical physicist, I have tried to
introduce such concepts either in the context of the simplest possible
examples, or in the context of a piece of well-known or important
piece of string theory lore.  A possible downside of this approach is
an occasional loss of clarity.

These lectures were constructed in with the help of some rather
standard references: the survey of differential geometry by Eguchi,
Gilkey, and Hansen \cite{egh}; some of the later chapters of the text
by Green, Schwarz, and Witten \cite{GSW}; appendix~B of Polchinski's
text \cite{Polch}; and the original papers by D.~Joyce on compact
manifolds of $G_2$ holonomy \cite{JoyceOne,JoyceTwo}.  The student of
string theory wishing to go beyond these lectures will find references
\cite{egh}-\cite{JoyceTwo} excellent jumping-off points.  Also, a set
of lectures on special holonomy from a pedagogical but more
mathematical point of view has appeared \cite{JoyceReview}.

\section{Lecture 1: on Calabi-Yau manifolds}
\label{CalabiYau}

\subsection{$A_k$ spaces}

The simplest non-trivial Calabi-Yau manifolds are four-dimensional,
even though the ones of primary interest in string model building are
six-dimensional.  To begin our acquaintance with four-dimensional
Calabi-Yau's, let's first consider some non-compact orbifolds.  In
particular, regard four-dimensional flat space as ${\bf C}^2$ (that
is, the Cartesian product of the complex plane with itself).  There is
a natural $SU(2)$ action on ${\bf C}^2$, where the two complex
coordinates form a doublet.  Let $\Gamma$ be a discrete subgroup of
$SU(2)$: for example, $\Gamma$ could be the group of transformations
acting on ${\bf C}^2$ like this:
  \eqn{CtwoAction}{
   {\bf Z}_{n+1}: \quad \pmatrix{ a \cr b } \to
    \pmatrix{ \omega & 0 \cr 0 & \omega^{-1} } \pmatrix{ a \cr b} \,,
  }
 where $(a,b)$ are coordinates on ${\bf C}^2$, and $\omega$ is any of
the $n+1$ complex numbers satisfying $\omega^{n+1}=1$.  The simplest
case would be $n=1$, so that $\Gamma = {\bf Z}_2$, and then the only
non-trivial transformation just changes the sign of $a$ and $b$---that
is, it reflects us through the origin of ${\bf R}^4 = {\bf C}^2$.  Now
form the orbifold ${\bf C}^2/\Gamma$.  Overlooking the singular point
at the origin, this is a manifold of holonomy $\Gamma$.  More
properly, we should call it an orbifold of holonomy $\Gamma$.

\begin{figure}[h]
   \centerline{\psfig{figure=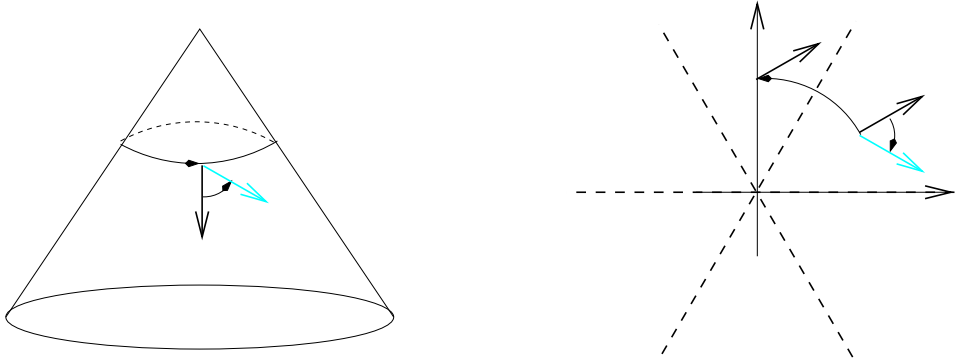,width=5in}}
   \caption{Left: parallel transport of a vector around the tip of a
cone changes its direction.  Right: the same parallel transport, where
the cone is thought of as a plane modded out by a discrete group.}
   \label{figA}
  \end{figure}

I haven't even defined holonomy yet, so how can we make such a
statement?  Consider a two-dimensional analogy: ${\bf R}^2$ admits a
natural $SO(2)$ action, and we could also embed $\Gamma = {\bf
Z}_{n+1} \subset SO(2) = U(1)$ in a natural way.  The orbifold ${\bf
R}^2/\Gamma$ is a cone of holonomy $\Gamma$.  This claim we can
understand just with pictures, and the complex case is only a slight
extension.  Suppose, as in figure~\ref{figA}, we take a vector at some
point away from the tip of the cone, and parallel translate it around
a loop.  This is easy to do in the original Cartesian coordinates on
${\bf R}^2$: the vector doesn't change directions.  For the loop that
I drew, and for $\Gamma = {\bf Z}_6$, the vector comes back to itself
rotated by an angle $\varphi = \pi/3$.
 This is what holonomy is all about: when vectors get
parallel-transported around some closed loop, their lengths remain
constant but their direction can change, and the holonomy group of an
n-dimensional real manifold is the subgroup of $O(n)$ that includes
all possible changes of direction for a vector so transported.  It is
a property of the manifold as a whole, not of any special point or
closed loop.  So for the example in figure~\ref{figA}, the holonomy
group is ${\bf Z}_6$, acting on the tangent plane of the orbifold in
the obvious way.  (We can define holonomy in the presence of an
orbifold singularity---or any other isolated singularity---just by
restricting to paths that avoid the singularity).  A generic, smooth,
orientable manifold has holonomy $SO(n)$.  The smaller the holonomy
group, the more special the manifold.  If the holonomy group is
trivial, the manifold is flat.  A non-vanishing Riemann tensor is a
local measure of non-vanishing holonomy, but we don't need to know
details of this yet.

The argument around figure~\ref{figA} can be repeated to show that
${\bf C}^2/{\bf Z}_{n+1}$ has holonomy ${\bf Z}_{n+1}$.  This orbifold
is called an $A_n$ singularity.  It's a singular limit of smooth
Calabi-Yau manifolds, as we'll see next.

The origin of ${\bf C}^2/{\bf Z}_{n+1}$ is a curvature singularity.  A
persistent theme in string theory is the resolution of singularities.
Singularity resolution is relatively easy work for Calabi-Yau
manifolds because we often have an algebraic description of them.  To
see how such descriptions arise, note that $a$ and $b$ are
double-valued on ${\bf C}^2/{\bf Z}_2$, but 
  \eqn{zs}{
   z_1 = a^2 \,,\quad z_2 = b^2 \,,\quad z_3 = ab
  }
 are single-valued.  We can pick any two of these as good local
coordinates for ${\bf C}^2/{\bf Z}_2$.  They are related by the
equation
  \eqn{ze}{
   z_3^2 = z_1 z_2 \,.
  }
 This is an {\it equation} for ${\bf C}^2/{\bf Z}_2$ in ${\bf C}^3$
(and the complex structure is correctly inherited from ${\bf C}^3$,
though the Kahler structure is not---if you don't know what this
means, ignore it for now).  A {\it nearby} submanifold of ${\bf C}^3$,
which is completely smooth, is
  \eqn{zdef}{
   z_3^2 - \epsilon^2 = z_1 z_2 \,,
  }
 or, after a linear complex change of variables
  \eqn{zdefTwo}{
   z_1^2 + z_2^2 + z_3^2 = \epsilon^2
  }
 where we can, without loss of generality, assume $\epsilon^2 \geq 0$.
Clearly, if $\epsilon=0$, we recover our original ${\bf C}^2/{\bf
Z}_2$ orbifold.

Writing $z_j = x_j + i y_j$, we can recast \zdefTwo\ as
  \eqn{zdefThree}{
   \vec{x}^2 - \vec{y}^2 = \epsilon^2 \,, \quad
    \vec{x} \cdot \vec{y} = 0 \,.
  }
 Now define $r^2 = \vec{x}^2 + \vec{y}^2 = \sum_{i=1}^3 |z_i|^2$.  For
large $r$, our deformed manifold, \zdef\ or \zdefTwo, asymptotically
approaches the original ``manifold,'' \ze.  Furthermore, we can easily
see that $r^2 \geq \epsilon^2$, and that for $r^2=\epsilon^2$, we have
to have $\vec{y}=0$ and $\vec{x}^2 = \epsilon^2$: this is a sphere of
radius $\epsilon$.  In fact, the manifold defined by \zdefTwo\ can
also be described as the cotangent bundle over $S^2$, denoted $T^*
S^2$.  To understand this, parametrize $S^2$ using a real vector
$\vec{w}$ with $\vec{w}^2 = \epsilon^2$.  Any 1-form on $S^2$ can be
expressed as $\vec{y} \cdot d\vec{w}$, where $\vec{y} \cdot \vec{w} =
0$.  The space of all possible 1-forms over a point on $S^2$ is ${\bf
R}^2$.  The total space of 1-forms over $S^2$, which we have called
$T^* S^2$, is thus some fibration of ${\bf R}^2$ over $S^2$.  And
we've just learned that this total space is parametrized by
$(\vec{w},\vec{y})$ with $\vec{w}^2=\epsilon^2$ and $\vec{y} \cdot
\vec{w} = 0$.  Now if we change variables from $\vec{w}$ to $\vec{x} =
\vec{w} \sqrt{1 + \vec{y}^2/\epsilon^2}$, we reproduce \zdefThree.

\begin{figure}[h]
  \centerline{\psfig{figure=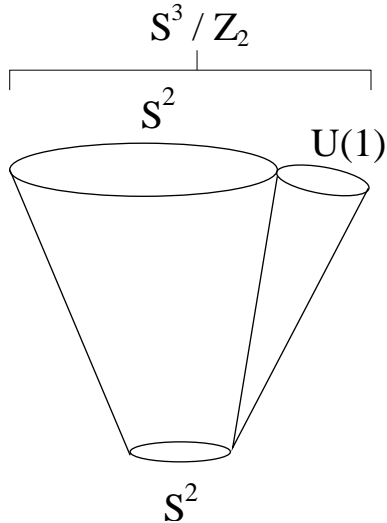,width=2in}}
   \caption{$S^3/{\bf Z}_2$ is a $U(1)$ fibration over $S^2$, and in
the interior, the $U(1)$ shrinks but the $S^2$ doesn't.}
   \label{figB}
  \end{figure}

Let's review what's happened so far.  The original orbifold, ${\bf
C}^2/{\bf Z}_2$, is a cone over $S^3/{\bf Z}_2$.  Note that $S^3/{\bf
Z}_2$ is smooth, because the ${\bf Z}_2$ action on $S^3$ induced from
\CtwoAction\ has no fixed points.  (It's the identification of
antipodal points).  In fact, $S^3/{\bf Z}_2$ is the $SO(3)$ group
manifold.  The higher $S^3/{\bf Z}_{n+1}$ are also smooth because the
${\bf Z}_{n+1}$ action has no fixed point on the $U(1)$ Hopf fiber.
Our algebraic resolution of the singularity led us to a smooth
manifold which was asymptotic to the cone over $S^3/{\bf Z}_2$, but
had a $S^2$ of radius $\epsilon$ at its ``tip'' rather than a
singularity.  This is illustrated schematically in figure~\ref{figB}.

This was just the beginning, because we have yet to really specify the
metric on the manifolds specified by \zdefTwo.  We should {\it not}
simply suppose that the metric naturally inherited from ${\bf C}^3$ is
the one we want.  In fact, the beautiful truth for these manifolds is
that there is a one-parameter family of Ricci-flat Kahler metrics
respecting the obvious $SO(3)$ symmetry of the equation \zdefTwo\
(explanation of the word ``Kahler'' will be forthcoming).  These
metrics have $SU(2)$ holonomy.  This means, precisely, that the spin
connection, $\omega_\mu{}^a{}_b$, generically an $SO(4)$ gauge field,
lies entirely in one $SU(2)$ subgroup of $SO(4) = SU(2)_L \times
SU(2)_R$.  By convention we could say that the holonomy group is
$SU(2)_L$.  Then a {\it constant} right-handed spinor field
$\epsilon_R$ obviously satisfies
  \eqn{nablaEps}{
   \nabla_\mu \epsilon = \partial_\mu \epsilon_R + 
     {1 \over 4} \omega_{\mu ab} \gamma^{ab} \epsilon_R = 0 \,,
  }
 just because the second term is a linear combination of the
generators of rotation in $SU(2)_L$, under which $\epsilon_R$ is
invariant.  The integrability condition of the equation \nablaEps\ is
  \eqn{nn}{
   [\nabla_\mu,\nabla_\nu] \epsilon_R = {1 \over 4} 
    R_{\mu\nu ab} \gamma^{ab} \epsilon_R = 0 \,,
  }
 for any $\epsilon_R$ such that $\gamma_5 \epsilon_R = -\epsilon_R$.
(That's an equivalent way of saying that a spinor is right-handed).
Thus, for {\it any} spinor $\epsilon$ (right-handed or not), 
  \eqn{RiemannReason}{\eqalign{
   R_{\mu\nu ab} \gamma^{ab} (1-\gamma_5) \epsilon &=
     R_{\mu\nu ab} \gamma^{ab} (1 - \gamma^1 \gamma^2 \gamma^3 \gamma^4)
      \epsilon  \cr
     &= R_{\mu\nu ab} \left( \gamma^{ab} - 
      {1 \over 2} \epsilon^{abcd} \gamma_{cd} \right) \epsilon  \cr
     &= \left( R_{\mu\nu ab} - {1 \over 2} \epsilon_{abcd}
      R_{\mu\nu}{}^{cd} \right) \gamma^{ab} \epsilon = 0 \,,
  }}
 and, evidently, this can be true if and only if the Riemann tensor is
self-dual:
  \eqn{SelfDual}{
   R_{\mu\nu ab} = {1 \over 2} \epsilon_{abcd} R_{\mu\nu}{}^{ab} \,.
  }
 Because \SelfDual\ looks a lot like the equations for an instanton in
non-abelian gauge theory, the metric of $SU(2)$ holonomy on \zdefTwo\
is known as a ``gravitational instanton.''  This metric is known
explicitly, and is called the Eguchi Hansen space, or $EH_2$:
  \eqn{EHTwo}{
   ds^2 = {dr^2 \over {1 - (\epsilon/r)^4}} + 
    r^2 \left( \sigma_x^2 + \sigma_y^2 + (1-(\epsilon/r)^4) \sigma_z^2
     \right) \,,
  }
 where
  \eqn{Sigmas}{\seqalign{\span\TC}{
   \sigma_x = \cos\psi d\theta + \sin\psi \sin\theta d\phi \quad
   \sigma_y = -\sin\psi d\theta + \cos\psi \sin\theta d\phi  \cr
   \sigma_z = d\psi + \cos\theta d\phi \,.
  }}
 It's worth noting that the metric on $S^3$ can be written as
  \eqn{SThree}{
   ds_{S^3}^2 = \sigma_x^2 + \sigma_y^2 + \sigma_z^2
    = d\theta^2 + \sin^2 \theta d\phi^2 + (d\psi + \cos\theta d\phi)^2 \,,
  }
 and the 1-forms $\sigma_i$ are invariant under the left action of
$SU(2)$ on $S^3 = SU(2)$.  To cover $S^3$ once, we should let $0 \leq
\theta \leq \pi$, $0 \leq \phi < 2\pi$, and $0 \leq \psi < 4\pi$.  On
the other hand, in the expression \EHTwo\ for the Eguchi-Hansen
metric, $\psi$ is restricted to range over $[0,2\pi)$.  Thus the
metric for large $r$ is indeed a cone over $SO(3) = S^3/{\bf Z}_2$:
the ${\bf Z}_2$ action on $S^3$ is just $\psi \to \psi+2\pi$.

Clearly, \EHTwo\ is the promised one-parameter family of metrics on
the resolved $A_1$ singularity.  The parameter is $\epsilon$, and one
can verify that the $S^2$ at $r=\epsilon$ indeed has radius
$\epsilon$ in the metric \EHTwo.  The $SO(3)$ symmetry of \zdefTwo\ is
included in the $SU(2)$ invariance of the $\sigma_i$.

Having thoroughly disposed of this simplest example of a special
holonomy metric, it's worth saying that a Calabi-Yau $n$-fold is, in
general, a manifold of $2n$ real dimensions whose holonomy group is
$SU(n)$ (or a subgroup thereof---but usually we mean that the holonomy
group is precisely $SU(n)$).  Any particle physicist will have
encountered the embedding of $SU(2)$ in $SO(4)$ as one of the
``chiral'' subgroups.  The inclusion of $SU(n)$ in $SO(2n)$ can be
described by saying that the $2n$ real-dimensional vector
representation of $SO(2n)$, which we could write as
$(x_1,y_1,x_2,y_2,\ldots,x_n,y_n)$, becomes the $n$-dimensional
complex representation of $SU(n)$, which we could write as
$(z_1,z_2,\ldots,z_n)$ where $z_j = x_j+iy_j$.  Having a holonomy
group $SU(n)$ necessarily means that the Calabi-Yau $n$-fold is
Ricci-flat: this is a frequently observed property of special holonomy
manifolds.  But not always: for instance, Kahler manifolds are $2n$
real-dimensional manifolds with holonomy group $U(n)$ (or a subgroup
thereof), and these aren't Ricci-flat unless the holonomy group is
contained in $SU(n)$.

The results described so far for the $A_1$ singularity admit
interesting generalizations in several directions:
 \begin{itemize}
 \item $A_n$ singularity: Here the natural, single-valued coordinates
are $z_1 = a^{n+1}$, $z_2 = b^{n+1}$, and $z_3 = ab$, and they are
related by the equation $z_3^{n+1}=z_1 z_2$, which can be deformed to
$\prod_{k=1}^{n+1} (z_3-\xi_k) = z_1 z_2$.  If the constants $\xi_k$
are all distinct, the deformed equation defines a smooth manifold in
${\bf C}^3$.  All such manifolds admit Ricci-flat metrics.  The ``tip
of the resolved cone'' is a rather more complicated geometry now:
there are $n(n+1)/2$ holomorphic embeddings of $S^2$ into a resolved
$A_n$ singularity, but only $n$ are distinct in homology.  Thus $b_2 =
n$ for these manifolds.
 \item $D_n$ and $E_6$, $E_7$, $E_8$ are the other finite subgroups of
$SU(2)$.  One can find algebraic descriptions and resolutions of ${\bf
C}^2/\Gamma$ for these cases as well, in a manner similar to the $A_n$
cases.
 \item Another important class of $SU(2)$ holonomy metrics is the
multi-center Taub-NUT solutions.  They are $U(1)$ fibrations over
${\bf R}^3$, with metric
  \eqn{dsTN}{\eqalign{
   &ds^2_{TN} = H d\vec{r}^{\,2} + H^{-1} (dx^{11} + 
    \vec{C} \cdot d\vec{r})^2 \qquad\hbox{where}  \cr
   &\nabla \times \vec{C} = -\nabla H \,,\quad
   H = \epsilon + {1 \over 2} \sum_{i=1}^{n+1} 
    {R \over |\vec{r}-\vec{r}_i|} \,.
  }}
 Clearly, $H$ is a harmonic function on ${\bf R}^3$.  There appears to
be a singularity in \dsTN\ when $\vec{r}$ is equal to one of the
$\vec{r}_i$, but in fact the manifold is completely smooth, for all
$\vec{r}_i$ distinct, provided $x^{11}$ is made periodic with period
$2\pi R$.  When $k>1$ of the $\vec{r}_i$ coincide, there is an
$A_{k-1}$ singularity.  An efficient way to see this is that, with $k$
of the $\vec{r}_i$ coincident, we've made the ``wrong'' choice of the
periodization of $x^{11}$: the right choice of period, to make the
local geometry non-singular, would have been $2\pi k R$.  We can get
from the right choice to the wrong choice by modding out $x^{11}$ by
${\bf Z}_k$, and now what's left is to convince yourself that this is
the same ${\bf Z}_k$ action that produced $A_{k-1}$ from ${\bf C}^2$.
If $\epsilon>0$, the geometry far for large $r$ is metrically the
product $S^1 \times {\bf R}^3$ (see figure~\ref{figC}).  If
$\epsilon>0$, asymptotically the space is a cone over $S^3/{\bf Z}_n$:
that is, in \dsTN\ with $\epsilon=0$ we have exhibited explicitly the
general metric of $SU(2)$ holonomy on a resolved $A_n$ singularity.
  \begin{figure}[h]
    \centerline{\psfig{figure=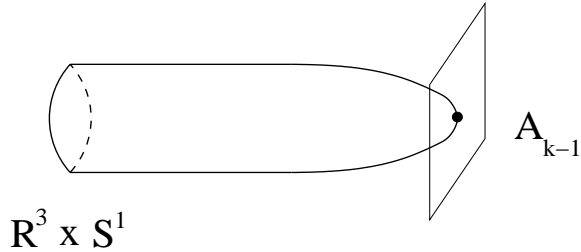,width=3in}}
   \caption{Single-center Taub-NUT ($k=1$ in \dsTN) interpolates
between ${\bf R}^3 \times S^1$ and an ${\bf R}^4$ which is
well-approximated by the tangent plane to the tip of the cigar.
Having $k$ centers coincident amounts to orbifolding by ${\bf Z}_k$ in
the $S^1$ direction, and results in an $A_{k-1}$ at the tip of the cigar.}
   \label{figC}
  \end{figure}

  \item It's now possible to outline the construction of a compact
Calabi-Yau 2-fold, also known as a K3 surface.  It's worth remarking
that all compact, smooth Calabi-Yau 2-folds with precisely $SU(2)$
holonomy are homeomorphic (not at all an obvious result).  Suppose we
start with $T^4 = {\bf R}^4/{\bf Z}^4$, where the lattice ${\bf Z}^4$
is just the one generated by the unit vectors $(1,0,0,0)$,
$(0,1,0,0)$, $(0,0,1,0)$, and $(0,0,0,1)$.  Now let us identify by the
action of ${\bf Z}_2$ which reflects through the origin: this is
precisely the ${\bf Z}_2$ action that we used to define the $A_1$
singularity, so evidently there will be such a singularity at the
origin.  Actually, on $T^4$ as a whole, there are $16$ fixed points of
the ${\bf Z}_2$ action, and each is an $A_1$ singularity: they are at
points $(r_1,r_2,r_3,r_4)$, where each $r_i$ can be chosen
independently as $0$ or $1/2$.  It's worth verifying that these are
all the fixed points.  A good way to go about it is to show that the
fixed points in ${\bf R}^4$ of the combined action of ${\bf Z}^4$ and
${\bf Z}_2$ are the images of the $16$ points we just mentioned under
action of the ${\bf Z}^4$.  A look at figure~\ref{figD}a) may help.
At any rate, we now have a compact but singular space, and its
holonomy is obviously ${\bf Z}_2$, with the usual caveat of avoiding
fixed points (the argument is the same as always: translate a vector
around the space, and the most it can do is switch its sign).  The
``Kummer construction'' of a smooth K3 space is to cut out a region of
radius $R$ around each of the $16$ $A_1$ singularities, and replace it
by a copy of the Eguchi-Hansen space, cut off at the same finite
radius $R$, and having an $S^2$ of radius $\epsilon>0$ at its tip.
This procedure works topologically because the surface $r=R$ of an
Eguchi-Hansen space is $S^3/{\bf Z}_2$, and that's the same space as
we got by cutting out a region around the $A_1$ singularity: the
boundary of $B^4/{\bf Z}_2$, where $B^4$ is a ball with boundary
$S^3$.  See figure~\ref{figD}b).  The metric does not quite match
after we've pasted in copies of $EH_2$, but it nearly matches: the
errors are $O(\epsilon^4/R^4)$.  Neglecting these small errors, we
have a smooth manifold of $SU(2)$ holonomy: the crucial point here is
that each $EH_2$ has the {\it same} $SU(2)$ subgroup of $SO(4)$ as its
holonomy group, namely the $SU(2)$ which contains the original
discrete ${\bf Z}_2$ holonomy of the $A_1$ singularity---and that
${\bf Z}_2$ is the same for all $16$ fixed points.  A {\it
non-trivial} mathematical analysis shows that the $O(\epsilon^4/R^4)$
can be smoothed out without enlarging the holonomy group.  It's easy
to understand from this analysis that K3 has $22$ homologically
distinct 2-cycles: $T^4$ started out with $6$ that are undisturbed by
the ${\bf Z}_2$ orbifolding (think of their cohomological partners,
for instance $dr_1 \wedge dr_2$, obviously ${\bf Z}_2$ even); and each
$EH_2$ adds one to the total because of the unshrunk $S^2$ at its tip.
As remarked earlier, all K3 surfaces are homeomorphic.  Hence all of
them have second Betti number $b_2 = 22$.
  \begin{figure}[h]
   \centerline{\psfig{figure=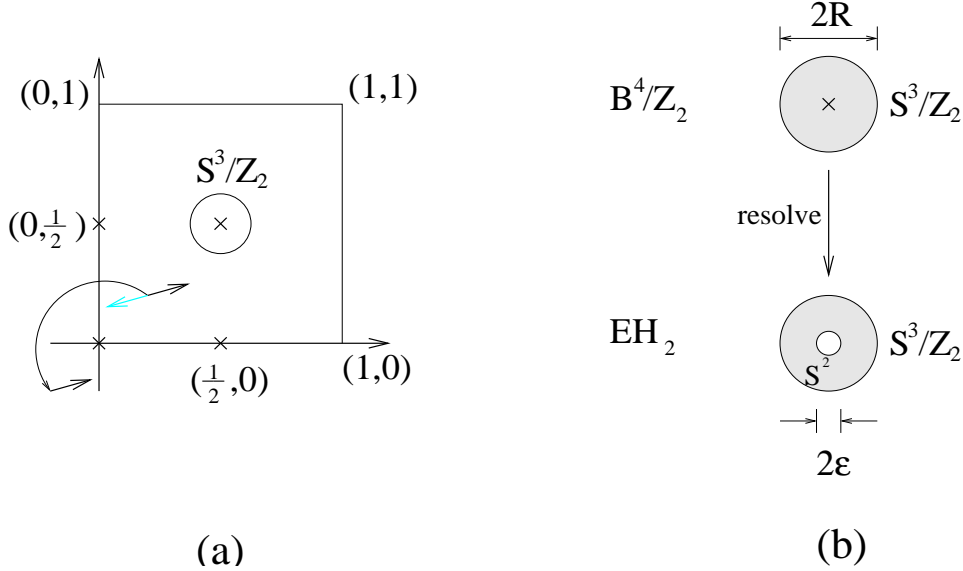,width=5in}}
   \caption{(a): Schematic description of $T^4/{\bf Z}_2$.  The unit
cell of a square torus is quotiented by the action of a ${\bf Z}_2$
whose fixed points are indicated by x's.  (Actually there would be
$2^4=16$ such fixed points for $T^4$, but we could only draw $T^2$
here).  Each fixed point is an $A_1$ singularity, so the boundary of a
region around it is $S^3/{\bf Z}_2$ in the quotient space.  The
quotient is an orbifold of ${\bf Z}_2$ holonomy: parallel transport of
a vector along a curve, plus its reflected image, are shown.  (b) We
resolve a $B^4/{\bf Z}_2$ region around each $A_1$ singularity into
the central portion of an Eguchi-Hansen space, with an unshrunk $S^2$
of radius $\epsilon$.}
   \label{figD}
  \end{figure}

\end{itemize}

It's worth reflecting for a moment on why we were able to get so far
in the study of the $A_k$ spaces just by manipulating complex
equations like $z_3^2 = z_1 z_2$.  This defining equation for the
$A_1$ space does not determine its metric, but it does determine its
complex structure.  That is, the notion of holomorphicity is inherited
from ${\bf C}^3$ to the subspace defined by the algebraic equation.
Another way to say it is we automatically have a distinguished way of
assembling four real coordinates into two complex coordinates.  Note
that we haven't said anything yet about the metric!  The natural
notion of a metric that is ``compatible'' with a given complex
structure is what's called a Kahler metric: it is one which can be
expressed locally as
  \eqn{KahlerDef}{
   ds^2 = 2 g_{i\bar{j}} dz^i d\bar{z}^{\bar{j}} \qquad\hbox{where}\quad
   g_{i\bar{j}} = {\partial^2 K \over \partial z^i \partial 
    \bar{z}^{\bar{j}}} \,,
  }
 for some function $K(z^i,\bar{z}^j)$ which is called the Kahler
potential.  Evidently, $K(z^i,\bar{z}^j)$ can be modified by the
addition of a holomorphic or an anti-holomorphic function.  It is
quite straightforward to show that the Christoffel connection
associated with a Kahler metric preserves the splitting of the tangent
plane into holomorphic and anti-holomorphic pieces: for instance, if a
vector points in the $z_1$ direction, then after parallel transport,
it may have components in the $z_1$ and $z_2$ directions, but none in
the $\bar{z}^1$ and $\bar{z}^2$ directions.  This is why Kahler
metrics on an $n$-complex-dimensional space necessarily have holonomy
$U(n)$.

Yau proved that if a smooth, compact manifold, admitting a complex
structure and a Kahler metric, obeys a certain topological condition
(vanishing of the first Chern class), then it's possible to find a
Ricci-flat Kahler metric.  (Some further facts are not so hard to
show: the Ricci-flat metric is unique given the cohomology class of
the Kahler form; and Ricci-flat Kahler metrics are precisely those
with holonomy contained in $SU(n)$).  By virtue of Yau's theorem, we
can go far in the study of $SU(n)$ holonomy manifolds just by
manipulating simple algebraic equations: the equations specify a
topology and a complex structure (inherited from the complex structure
in the flat space or projective space in which we write the defining
equations) and provided we can demonstrate the (rather weak)
topological hypotheses of Yau's theorem, we can be sure of the
existence of a $SU(n)$ holonomy metric even if we can't write it down.
Perhaps the simplest way to look at it is that you get to $U(N)$
holonomy just by knowing the complex structure.  The Kahler metric is
detailed and difficult information, but a lot of interesting facts can
be learned without knowing much about it other than its existence.

We have discussed some of the simplest special holonomy manifolds, and
sketched the Kummer construction for a compact K3; but much much more
remains unsaid.  There are highly developed ways of constructing
Calabi-Yau three-folds, of which elliptic fibration, toric geometry,
and the intersection of algebraic varieties in complex projective
spaces deserve special mention.  Far too much is in the literature to
even summarize here; but the interested reader will find much already
in the references to these lectures.

\subsection{Non-linear sigma models and applications to string theory}
\label{Sigma}

I find it irresistible at this point to detour into some applications
of notions from special holonomy to supersymmetry and string theory.
In four dimensions, the most general renormalizable lagrangian for a
single chiral superfield, $\Phi = \phi + \theta^\alpha \psi_\alpha +
\theta^\alpha \theta_\alpha F$, is
  \eqn{ChiralL}{
   {\cal L} = \int d^4 \theta \, \Phi^\dagger \Phi + 
    \left( \int d^2 \theta \, W(\Phi) + h.c. \right) \,,
  }
 with $W(\Phi)$ some cubic polynomial.  Let us work in Euclidean
signature.  The most general {\it effective} action for several chiral
superfields (that is, a totally general local form up to two
derivative) is the following:
  \eqn{LChiral}{\eqalign{
   {\cal L} &= \int d^4 \theta K(\Phi_i,\Phi_i^\dagger) + 
    \left( \int d^2 \theta W(\Phi) + h.c. \right)  \cr
    &= g_{i\bar{j}} \partial_\mu \phi^i \partial^\mu
     \bar\phi^{\bar{j}} + g_{i\bar{j}} \psi^i \slashed\partial
      \bar\psi^{\bar{j}} + 
      g^{i\bar{j}} {\partial W \over \partial\phi^i}
       {\partial \bar{W} \over \partial \bar\phi^{\bar{j}}} + 
       \left( {\partial^2 W \over \partial \phi^i \partial \phi^j}
        \psi^i \psi^j + h.c. \right) + \ldots \,,
  }}
 where in the last line I have eliminated the auxiliary fields $F_i$
through their algebraic equations of motion.  In expanding things out
in components I have left out various interaction terms, and I have
not been particularly careful with all factors of $2$ and signs.

If the superpotential is $0$, then the lagrangian is just
  \eqn{NLsM}{
   L = g_{i\bar{j}}(\phi^k,\bar\phi^{\bar{k}})
    \partial_\mu \phi^i \partial^\mu \bar\phi^{\bar{j}} + \hbox{fermions}
     \,,
  }
 which is just a non-linear sigma model with a Kahler target.  There
are various reasons to be interested in the lagrangians \LChiral\ and
\NLsM, but let us point out one that is particularly relevant to
string theory.  If we make a dimensional reduction to two dimensions,
setting $\phi^i = Z^i/\sqrt{2\pi\alpha'}$, then we obtain an action
  \eqn{STwo}{
   S = {1 \over 2\pi\alpha'} \int d^2 z \, 
     g_{i\bar{j}}(Z^k,\bar{Z}^{\bar{k}})
    \left( \partial_z Z^i \partial_{\bar{z}} \bar{Z}^{\bar{j}} + 
     \partial_{\bar{z}} Z^i \partial_z \bar{Z}^{\bar{j}} + \hbox{fermions}
      \right) \,.
  }
 The bosonic part written out explicitly is precisely the so-called
Polyakov action, $S_{\rm Pol} = {1 \over 2\pi\alpha'} \int d^2 z \,
g_{ab} \partial_z X^a \partial_{\bar{z}} X^b$, written in terms of complex
variables, $Z^j \propto X^{2j-1} + i X^{2j}$.  The action \STwo\
describes strings propagating on a Kahler manifold.  We know (see for
instance E.~D'Hoker's lectures at this school) that conformal
invariance forces this manifold to be ten-dimensional and {\it Ricci
flat}, in the leading approximation where $\alpha'$ is small compared
to characteristic sizes of the manifold.  For instance, the target
space could be a Calabi-Yau manifold times flat space: this is part of
the standard strategy for getting four-dimensional models out of the
heterotic string (more on this later).  

A simpler example would be for the target space just to be ${\bf R}^6$
times the Eguchi-Hansen space, $EH_2$.  (In fact, we could even use
the singular orbifold limit, provided $\int_{S^2} B_2 = \pi$; but it
is too much to consider here in detail how string physics can be
smooth on a singular geometry).  Pursuing our simple ${\bf R}^6 \times
EH_2$ example a little further: an obvious thing for a string to do is
to wrap the $S^2$ in $EH_2$.  The string is then an instanton with
respect to the ${\bf R}^6$ directions, and to compute its contribution
to the path integral, the first thing we have to know is the minimal
classical action for such a string.\footnote{We would eventually have
in mind formulating a string theory in ${\bf R}^{5,1}$ via Wick
rotation from ${\bf R}^6$---or, in the more physically interesting
case of a Calabi-Yau three-fold, in ${\bf R}^{3,1}$ via Wick rotation
from ${\bf R}^4$---but we carry on in the hallowed tradition of doing
all computations in Euclidean signature until the very end.}  To this
end, it is worth recalling that the Polyakov action coincides with the
Nambu-Goto action after the worldsheet metric is eliminated through
its algebraic equation of motion.  So the minimal action will be
attained by a worldsheet wrapped on the minimal area $S^2$.  Finding
this $S^2$ is straightforward work, since we have the explicit metric
for $EH_2$: it's obviously $r=\epsilon$.  But for a more general
discussion, it's worth introducing a little more technology, in the
form of the Kahler form
  \eqn{JDef}{
   J = i g_{i\bar{j}}(Z^k,Z^{\bar{k}}) dZ^i \wedge d\bar{Z}^{\bar{j}} \,.
  }
 Since both $g_{i\bar{j}} \partial_z Z^i \partial_{\bar{z}}
\bar{Z}^{\bar{j}}$ and $g_{i\bar{j}} \partial_{\bar{z}} Z^i \partial_z
\bar{Z}^{\bar{j}}$ are everywhere positive quantities, it's clear that
  \eqn{JIneq}{
   {1 \over 2\pi\alpha'} \int_{S^2} J = 
    {1 \over 2\pi\alpha'} \int_{S^2} d^2 z \, g_{i\bar{j}}
     \left( \partial_z Z^i \partial_{\bar{z}} \bar{Z}^{\bar{j}} - 
      \partial_{\bar{z}} Z^i \partial_z \bar{Z}^{\bar{j}} \right) \leq
    S_{\rm Pol}
  }
 with equality precisely if $g_{i\bar{j}} \partial_{\bar{z}} Z^i
\partial_z \bar{Z}^{\bar{j}} = 0$, which is equivalent to
$\partial_{\bar{z}} Z^i = 0$ for all $i$.  This last equation
expresses the condition that the map $z \to Z^i(z)$ is a {\it
holomorphic embedding} of the worldsheet into the target spacetime.
Obviously, we could consider anti-holomorphic embeddings, and prove in
an analogous way that precisely they saturate the inequality ${1 \over
2\pi\alpha'} \int_{S^2} J \geq -S_{\rm Pol}$.  A string
anti-holomorphically embedded in $EH_2$ would just be one at
$r=\epsilon$, wrapping the $S^2$ with the opposite orientation.  Thus
we have world-sheet instantons and world-sheet anti-instantons.

The inequality \JIneq\ is deceptively simple.  Actually it illustrates
a very powerful notion: calibration.  To see things in a properly
general light, first note that we didn't need the two-cycle to be
$S^2$: it could have been any homologically non-trivial two-cycle,
call it $\Sigma$.  Furthermore, we could have derived a pointwise form
of the inequality in \JIneq\ (obvious since we didn't need any
integrations by parts to get the inequality we did derive).  That
pointwise form would say that the pullback of the Kahler form $J$ to
the worldsheet is equal to a multiple of the volume form (defined
through the induced metric on the worldsheet), and the multiple is a
function that never exceeds $1$.  A final important ingredient to the
setup of a calibration is that $J$ is closed, $dJ = 0$.  This arises
because $J = i\partial\bar\partial K$, where $\partial$ is the exterior
derivative with respect to the $Z^i$'s, and $\bar\partial$ is the
exterior derivative with respect to the $\bar{Z}^{\bar{i}}$'s.  So to
state the whole setup once and for all and with full generality: a
calibration is a closed $p$-form which restricts (or, more precisely,
pulls back) onto any $p$-submanifold to a scalar multiple of the
induced volume form, where the multiple is nowhere greater than $1$;
and a calibrated cycle is one whose induced volume form precisely
coincides with the pullback of the calibration form.  An inequality
like \JIneq\ then ensures that the volume of the calibrated cycle is
minimal among all possible cycles in its homology class: this is
because the integral of the calibrating form (i.e.{} the left hand
side of \JIneq) depends only on the homology of what you're
integrating it over.

Suppose now we have a compactification of string theory from ten
dimensions to four on a (compact) Calabi-Yau three-fold, $CY_3$.  If
we pick a basis $N^A$ of homology two-cycles for $CY_3$, then we could
define the Kahler parameters as $v^A = \int_{N^A} J$.  From the
preceding discussion, $v^A$ is just the minimal area two-cycle in a
given equivalence class.  A natural complexification of $v^A$ is
  \eqn{TDef}{
   T^A = \int_{N^A} (J + iB) \,,
  }
 where $B$ is the NS 2-form, assumed to have $dB = 0$.  The $T^A$ are
the so-called complexified Kahler moduli of the Calabi-Yau
compactification.  The claim is that they become massless complex
fields in four dimensions.  To see this in precise detail, we should
perform a rigorous Kaluza-Klein reduction.  Without going that far, we
can convince ourselves of the claim by expanding
  \eqn{JBExpand}{
   J+iB = \sum_A T^A \omega_A \,,
  }
 where the $\omega_A$ are harmonic two-forms with $\int_{N^A} \omega_B
= \delta^A_B$; and \JBExpand\ is basically the beginnings of a
Kaluza-Klein reduction, where $T^A$ depends only on the four
non-compact dimensions.  Since the left hand side of \JBExpand\ is
harmonic (or may at least be made so by a gauge choice) and the
$\omega_A$ are harmonic, the $T^A$ are indeed massless fields in four
dimensions.  Compactification on $CY_3$ preserves $1/4$ of
supersymmetry (a theme to be developed more systematically in the next
lecture), which means ${\cal N}=1$ supersymmetry in $d=4$ for a
heterotic string compactification, and ${\cal N}=2$ supersymmetry in
$d=4$ for a type II string compactification.  Since we have at least
${\cal N}=1$ supersymmetry, the complex scalar fields $T^A$ must be
components of chiral superfields, with an action of the form \ChiralL,
for some Kahler target manifold that describes all possible values of
the complexified Kahler moduli for a given Calabi-Yau
compactification.\footnote{We have not substantially constrained how
these moduli might couple to other sorts of matter.  This issue is
beyond the scope of the present lectures.}  What a mouthful!  Now
comes the nice part: having learned that the $T^A$ are massless fields
based on an argument that applied for {\it any} Calabi-Yau, we can
confidently say that $V=0$ identically, so also the superpotential
$W=0$.  These are classical statements, because the argument that the
$T^A$ were massless was based on classical field equations.  However,
as is often the case, $W$ is protected against contributions from
loops by the unbroken ${\cal N}=1$ supersymmetry.  More precisely, a
Peccei-Quinn symmetry for $\int_{N^A} B$, plus holomorphy, protects
$W$ against {\it all} perturbative string corrections.  There are in
fact non-perturbative corrections that come from the world-sheet
instantons discussed above: the action of such an instanton is
  \eqn{WIAction}{
   S = {1 \over 2\pi\alpha'} \int_{N^A} (J+iB) = {T^A \over 2\pi\alpha'} \,,
  }
 and because of the explicit $T^A$-dependence, we obviously must
expect some nonperturbative $e^{-T^A/2\pi\alpha'}$ contribution to $W$
to arise from these instantons.

This is about all one can learn about the dynamics of complexified
Kahler moduli for $CY_3$ compactifications of superstrings based on
${\cal N}=1$ supersymmetry.  It's actually quite a lot: we have
non-linear sigma model dynamics on a Kahler manifold whose complex
dimension is $b_2$ of the $CY_3$, corrected only non-perturbatively in
the small dimensionless parameters $T^A/2\pi\alpha'$.  More can be
learned, however, if there is ${\cal N}=2$ supersymmetry---that is,
for $CY_3$ compactifications of a type II superstring.  Then one can
show that the Kahler metric on the moduli space follows from the
Kahler potential
  \eqn{KNTwo}{
   K = -\log {\cal W}(\Re T^A) \qquad
   {\cal W} = \int_{CY_3} J \wedge J \wedge J \,,
  }
 where $J$ is the Kahler form of the $CY_3$ (but $K$ is the Kahler
potential for the many-dimensional moduli space, and as such is a
function of $T^A$ and $\bar{T}^{\bar{A}}$).  Explaining how \KNTwo\
arises from ${\cal N}=2$ supersymmetry would take us too far afield;
it is enough for us to know that, whereas ${\cal N}=1$ supersymmetry
usually protects only the holomorphic object $W$ from corrections,
${\cal N}=2$ supersymmetry tightly constrains the Kahler form as well,
protecting it in this case from all perturbative string corrections.
There are worldsheet instanton corrections, as before.

A substantial omission in our treatment is that we haven't discussed
complex structure moduli.  Understanding them, and also the worldsheet
origin of both types of moduli, is crucial to the formulation of
mirror symmetry in string theory.  The reader may wish to consult
TASI lectures from previous years (for instance \cite{GreeneTASI}) for
an introduction to these fascinating topics.

A truly remarkable property of heterotic string theory dynamics is
that the form \KNTwo\ continues to hold true, modulo similar
non-perturbative corrections, in ${\cal N}=1$ compactifications of the
heterotic string with the ``standard embedding'' of the spin
connection in the gauge group.  ``Standard embedding'' means that one
sets gauge potentials $A_\mu{}^I{}_J$ in a particular $SU(3)$ subgroup
of $SO(32)$, or of $E_8 \times E_8$, equal to the spin connection
$\omega_\mu{}^a_b$ of the $CY_3$.  In contrast to the results
presented so far, the fact that \KNTwo\ persists for these ${\cal
N}=1$ constructions goes beyond anything one could understand based
only on low-energy effective field theory, and is truly stringy in its
origin.  Before returning to the narrower venue of special holonomy,
let us then detour into a demonstration of this claim.  Amusingly,
almost all the tools we will use have already been introduced.

The basic point is that, for the standard embedding, the $CY_3$ part
of the heterotic worldsheet CFT is identical to the corresponding part
of the type II worldsheet CFT.  Because the heterotic CFT factorizes
into a ${\bf R}^{3,1}$ part, a $CY_3$ part (to be described), and an
``extra junk'' part, the physical dynamics of the $CY_3$ is the same
for the heterotic and type II constructions.  It's as if there were a
``secret'' ${\cal N}=2$ supersymmetry in the heterotic string.  To
write down type II superstring propagation on a $CY_3$, we need to
make the non-linear sigma model \STwo\ explicitly supersymmetric.
With the help of superfields
  \eqn{SuperXDef}{
   {\cal X}^a = X^a + i\theta \psi^a + i\bar\theta \tilde\psi^a + 
    \theta \bar\theta F^a \,,
  }
 one can write a simple supercovariant worldsheet action:
  \eqn{SuperS}{\eqalign{
   S_{CY_3} &= {1 \over 2\pi\alpha'} \int d^2 z d^2 \theta \,
     g_{ab}({\cal X}) D_{\bar\theta} {\cal X}^a D_\theta {\cal X}^b  \cr
    &= {1 \over 2\pi\alpha'} \int d^2 z \Bigg[
     g_{ab}(X) \partial X^a \bar\partial X^b + g_{ab}
      (\psi^a D_{\bar{z}} \psi^b + \tilde\psi^a D_z \tilde\psi^b)
     \cr &\qquad\quad{} + {1 \over 2} R_{\mu\nu\rho\sigma}(X) 
      \psi^\mu \psi^\nu \tilde\psi^\rho \tilde\psi^\sigma \Bigg] \,,
  }}
 where the second equality holds after auxiliary fields have been
algebraically eliminated.  The covariant derivatives are defined as
follows:
  \eqn{SuperDDef}{\eqalign{
   & D_\theta = \partial_\theta + \theta \partial_z \qquad
   D_{\bar\theta} = \partial_{\bar\theta} + \bar\theta \partial_z  \cr
   & D_{\bar{z}} \psi^a = \partial_{\bar{z}} \psi^a + 
    \partial_{\bar{z}} X^b \Gamma^a_{bc}(X) \psi^c  \cr
   & D_z \tilde\psi^a = \partial_z \tilde\psi^a + 
    \partial_z X^b \Gamma^a_{bc}(X) \tilde\psi^c \,.
  }}
 The complicated second term in $D_{\bar{z}} \psi^a$ and $D_z
\tilde\psi^a$ are the pull-backs of the Calabi-Yau connection to the
string worldsheet.  The full action for type II superstrings on ${\bf
R}^{3,1} \times CY_3$ is 
  \eqn{SFullII}{
   S_{II} = {1 \over 2\pi\alpha'} \int d^2 z d^2 \theta \, 
    \eta_{\mu\nu} D_{\bar\theta} {\cal X}^\mu D_\theta {\cal X}^\nu + 
    S_{CY_3} \,.
  }
 The heterotic string possesses only the anti-holomorphic fermions
$\tilde\psi^M$: instead of the corresponding ten holomorphic fermions
$\psi^M$, the heterotic string has $32$ holomorphic fermions
$\lambda^I$.  (The choice of GSO projection determines whether we have
$SO(32)$ or $E_8 \times E_8$ as the gauge group.  In the latter case,
$SO(16) \times SO(16)$ is manifest in the above description, as
rotations of the $\lambda^I$'s in two sets of 16.  For further details
about the heterotic string, standard string theory texts should be
consulted).  The action of the heterotic string is
  \eqn{SHet}{
   S_{Het} = {1 \over 2\pi\alpha'} \int d^2 z \left[
    g_{MN} \partial X^M \bar\partial X^N + 
    g_{MN} \tilde\psi^M D_z \tilde\psi^N + 
    \delta_{IJ} \lambda^I {\cal D}_{\bar{z}} \lambda^J + 
    {1 \over 2} F_{MN}^{IJ} \lambda^I 
    \lambda^J \tilde\psi^M \tilde\psi^N \right]
   }
 where the only new derivative we need to define is
  \eqn{DCalDef}{
   {\cal D}_{\bar{z}} \lambda^I = \partial_z \lambda^I + 
    A_M^{IJ}(X) \partial_{\bar{z}} X^M \lambda^J \,,
  }
 the second term being the heterotic gauge field pulled back to the
worldsheet.  (It's easiest to think of the $A_M^{IJ}$ either as
$SO(32)$ gauge fields, or in the $E_8 \times E_8$ case as $SO(16)
\times SO(16)$ gauge fields, which have to be augmented by some other
fields to make up the full $E_8 \times E_8$, but these other fields
will never be turned on in our construction).  Now for the punch-line:
we can embed $S_{CY_3}$ into $S_{Het}$ by ``borrowing'' six of the
$\lambda^I$ to replace the six lost $\psi^a$.  More explicitly,
  \eqn{PsiReplace}{\eqalign{
   & \psi^I \equiv e^I_a(X) \psi^a \to \lambda^I \,,\quad
   \omega_a^{IJ} \to A_a^{IJ} \,,\quad
     R_{ab}{}^{IJ} \to F_{ab}{}^{IJ}  \cr
   & \delta_{IJ} e^I_a e^J_b = g_{ab}  \quad I = 1,\ldots,6 \qquad
   D_{\bar{z}} \psi^I = \partial_{\bar{z}} \psi^I + 
    \partial_{\bar{z}} X^a \omega_a^{IJ}(X) \psi^J \,.
  }}
 Thus, quite literally, we are embedding a particular $SU(3) \subset
SU(4) = SO(6) \subset SO(16)$, and the $SO(16)$ is either part of
$SO(32)$ or $E_8$.  Clearly, $SU(3) \subset SU(4)$ in only one way,
and $SO(6) \subset SO(16)$ so that $SO(6)$ rotates only $6$ components
of the real vector representation of $SO(16)$).

A lesson to remember, even if not all the details registered, is that
the spin connection can be thought of as just another connection
(acting on the tangent bundle so that $\nabla_a v^I = \partial_a v^I +
\omega_a{}^I{}_J v^J$), and it is not only well-defined, but in fact
quite convenient, to set some of the gauge fields of the heterotic
string equal to the spin connection of $SU(3)$ holonomy that we know
exists on any Calabi-Yau.  Less minimal choices have been extensively
explored---see for example D.~Waldram's lectures at this school
\cite{WaldramTASI}.

\section{Lecture 2: on $G_2$ holonomy manifolds}
\label{GTwo}

Given that all string theories can be thought of as deriving from a
single eleven-dimensional theory, M-theory, by a chain of dualities,
it is natural to ask what are the sorts of seven-dimensional manifolds
we can compactify M-theory on to obtain minimal supersymmetry in
four-dimensions.\footnote{Some people might prefer the phrasing, ``All
string theories are special limits of a mysterious theory, M-theory,
of which another limit is eleven-dimensional supergravity.''  I will
prefer to use M-theory in its more restrictive sense as a theory
emphatically tied to eleven dimensions---in other words, the as-yet
unknown quantum completion of eleven-dimensional supergravity.}  This
is the most obvious string theory motivation for studying
seven-manifolds of $G_2$ holonomy, as indeed we shall see that
M-theory on such manifolds leads to ${\cal N}=1$ supersymmetry in
$d=4$.

\begin{figure}[h]
   \centerline{\psfig{figure=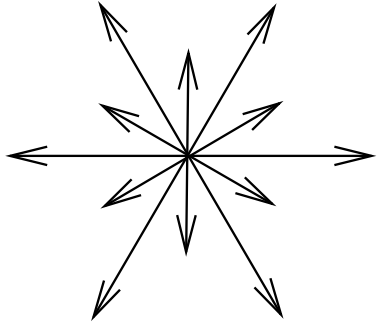,width=2in}}
   \caption{The Dynkin diagram for $G_2$.  The weights comprising the
${\bf 7}$ are the six short roots plus one node at the origin.}
   \label{figE}
  \end{figure}

But what is $G_2$?  It can be defined as the subgroup of $SO(7)$ whose
action on ${\bf R}^7$ preserves the form
  \eqn{PhiLocal}{\eqalign{
   \varphi &= dy^1 \wedge dy^2 \wedge dy^3 + 
      dy^1 \wedge dy^4 \wedge dy^5 + 
      dy^1 \wedge dy^6 \wedge dy^7 + 
      dy^2 \wedge dy^4 \wedge dy^6  \cr
     &\qquad{} - dy^2 \wedge dy^5 \wedge dy^7 -
      dy^3 \wedge dy^4 \wedge dy^7 -
      dy^3 \wedge dy^5 \wedge dy^6  \cr
   &\equiv {1 \over 6} \varphi_{abc} dy^a dy^b dy^c \,.
  }}
 The $\varphi_{abc}$ happen to be the structure constants for the
imaginary octonions.  We will not use this fact, but instead take the
above as our {\it definition} of $G_2$.  Let's now do a little group
theory.  $SO(7)$ has rank 3 and dimension 21.  Three obvious
representations are the vector ${\bf 7}$, the spinor ${\bf 8}$, and
the adjoint ${\bf 21}$.  $G_2$, on the other hand, has rank 2 and
dimension 14.  See figure~\ref{figE}.  

 It has two obvious representations: the fundamental ${\bf 7}$
(comprising the short roots plus one weight at the origin) and the
adjoint ${\bf 14}$. As a historical note, it's worth mentioning that
$G_2$ enjoyed brief popularity as a possible group to describe flavor
physics: the ${\bf 7}$ was supposed to be the multiplet of
pseudoscalar mesons.  That looked OK until it was realized that the
$\eta$ had to be included in this multiplet, which made the ${\bf 8}$
of $SU(3)$ clearly superior.  Besides, spin $3/2$ baryons almost
filled out the ${\bf 10}$ of $SU(3)$, and then the discovery of the
$\Omega^-$ completed that multiplet and clinched $SU(3)$'s victory.
To return to basic group theory, it's worth noting some branching
rules:
  \eqn{Branch}{\seqalign{\span\TL & \span\TR \qquad\qquad & \span\TL & \span\TR}{
   SO(7) &\supset G_2               & G_2 &\supset SU(3)  \cr
   {\bf 21} &= {\bf 14} 
    \oplus {\bf 7}                  & {\bf 7} &\supset {\bf 3}  
                                       \oplus \bar{\bf 3} \oplus {\bf 1}  \cr
   {\bf 7} &= {\bf 7}               & {\bf 14} &= {\bf 8}_{\rm adj} \oplus
                                     {\bf 3} \oplus \bar{\bf 3}  \cr
   {\bf 8} &= {\bf 7} \oplus {\bf 1}
  }}
 The second rule in the right column suggests another construction of
$G_2$, as $SU(3)$ plus generators in the ${\bf 3}$ and the $\bar{\bf
3}$---this is similar to the construction of $E_8$ from $SO(16)$ plus
spinor generators.

\begin{figure}[h]
   \centerline{\psfig{figure=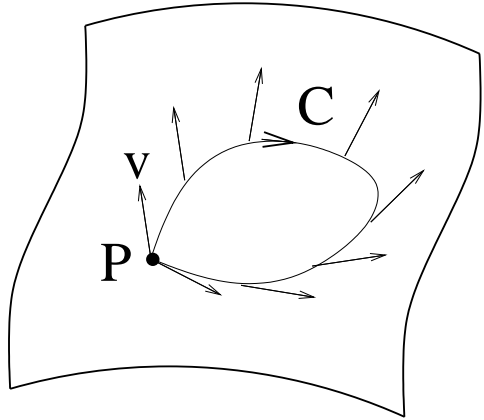,width=2.5in}}
   \caption{Parallel transport of a vector $v$ around a curve $C$.
Upon returning to the point of origin $P$, $v$ has undergone some
rotation, which for a seven-manifold is an element of $SO(7)$.}
   \label{figF}
  \end{figure}

The construction of $G_2$ as a subgroup of $SO(7)$ makes it clear that
$G_2$ is a possible holonomy group of seven-manifolds.  Before
explaining this in detail, let us re-orient the reader on the concept
of holonomy.  Recall that on a generic seven-manifold, parallel
transport of a vector around a closed curve brings it back not to
itself, but to the image of itself under an $SO(7)$ transformation
which depends on the curve one chooses.  See figure~\ref{figF}.
 The reason that the transformation is in $SO(7)$ is that the {\it
length} of the vector is preserved: parallel transport means $t^\mu
\nabla_\mu v^\alpha = 0$ along the curve $C$, and this implies $t^\mu
\nabla_\mu (g_{\alpha\beta} v^\alpha v^\beta) = 0$ (because
$\nabla_\mu g_{\alpha\beta} = 0$); so indeed the length of the vector
$v$ is the same, all the way around the curve.  Suppose we now choose
some seven-bein $e^a_\mu$, satisfying $\delta_{ab} e^a_\alpha e^b_\beta =
g_{\alpha\beta}$.  Parallel transporting all seven of these 1-forms
around our closed curve $C$ results in
  \eqn{Oe}{
   e^a_\alpha \to O^a{}_b e^b_\mu \,,
  }
 where $O^a{}_b \in SO(7)$.  Parallel transport in this context means
transport with respect to the covariant derivative $\nabla_\nu e^a_\mu
= \partial_\nu e^a_\mu - \Gamma^\rho_{\nu\mu} e^a_\rho$: that is, we
treat $a$ merely as a label.  One often defines another covariant
derivative, $D_\mu$, such that a flat index $a$ results in an extra
term involving the spin connection: thus for instance
  \eqn{SpinConnect}{
   D_\nu e^a_\mu = \partial_\nu e^a_\nu - \Gamma^\rho_{\nu\mu} e^a_\rho
     + \omega_\nu{}^a{}_b e^b_\mu \,.
  }
 The spin connection can then be defined by the equation $D_\nu
e^a_\mu = 0$.  

Thus far our setup has nothing to do with $G_2$: we have merely
explained (or re-explained) some standard aspects of differential
geometry.  Now suppose our seven-manifold is special, in that for {\it
some} choice of seven-bein $e^a_\mu$, the three-form
  \eqn{PhiManifold}{\eqalign{
   \varphi &= e^1 \wedge e^2 \wedge e^3 + 
      e^1 \wedge e^4 \wedge e^5 + 
      e^1 \wedge e^6 \wedge e^7 + 
      e^2 \wedge e^4 \wedge e^6  \cr
     &\qquad{} - e^2 \wedge e^5 \wedge e^7 -
      e^3 \wedge e^4 \wedge e^7 -
      e^3 \wedge e^5 \wedge e^6  \cr
   &\equiv {1 \over 6} \varphi_{abc} e^a e^b e^c
  }}
 satisfies $\nabla_\mu \varphi_{\alpha\beta\gamma} = 0$.  That means,
in particular, that if we parallel transport $\varphi$ around $C$, it
comes back to itself.  Rephrasing this statement using \Oe\ and the
concise form $\varphi = {1 \over 6} \varphi_{abc} e^a e^b e^c$, we see
that $\varphi_{abc} O^a{}_d O^b{}_e O^c{}_f = \varphi_{def}$.  So the
$SO(7)$ transformation $O^a{}_b$ is actually an element of $G_2$; and
since the curve $C$ was arbitrary, the manifold's holonomy group is
$G_2$.

The presentation of the previous paragraph is in the order that my
intuition suggests; however it's actually backwards according to a
certain logic.  A mathematician might prefer to state it this way: it
so happens that preservation of the form \PhiLocal\ under a general
linear transformation {\it implies} preservation of the metric
$\delta_{ab}$.  So we could start with a manifold $M_7$ endowed only
with differential structure, choose a globally defined three-form
$\varphi$ on it, {\it determine} the metric $g_{\mu\nu}$ in terms of
$\varphi$,\footnote{A formula for the metric in terms of $\varphi$
will be given in section~\ref{DSix}.  The validity of this formula
already requires that $\varphi$ have some non-degeneracy properties.
A more careful analysis can be found in \cite{hitchin}.}  determine
the connection $\nabla_\mu$ in terms of $g_{\mu\nu}$, and then ask
that $\nabla_\mu \varphi_{\alpha\beta\gamma} = 0$ in order to have a
$G_2$ holonomy manifold.  This amounts to a set of hugely non-linear
differential equations for the three-form coefficients
$\varphi_{\alpha\beta\gamma}$.

The decomposition ${\bf 8} = {\bf 7} \oplus {\bf 1}$ of the spinor of
$SO(7)$ into representations of $G_2$ is important, because it means
that $G_2$ holonomy manifolds admit precisely one covariantly constant
spinor.  To construct it, start at any point $P$, choose $\epsilon$ at
$P$ as the singlet spinor according to the above decomposition, and
then parallel transport $\epsilon$ everywhere over the manifold.
There is no path ambiguity because the spinor always stays in the
singlet representation of $G_2$.  All other spinors are shuffled
around by the holonomy: only the one we have constructed satisfies
$\nabla_\mu \epsilon = 0$.  The equation for preserved supersymmetry
in eleven-dimensional supergravity, with the four-form $G_{(4)}$ set
to zero, is
  \eqn{PreserveSUSY}{
   \delta\psi_\mu = \nabla_\mu \eta = 0 \,.
  }
 For an eleven-dimensional geometry ${\bf R}^{3,1} \times M_7$, where
$M_7$ has $G_2$ holonomy, the solutions for $\eta$ in \PreserveSUSY\
are precisely $\epsilon$ tensored with a spinor in ${\bf R}^{3,1}$:
that is, compactification on $M_7$ preserves one eighth of the
possible supersymmetry, which amounts to ${\cal N}=1$ in $d=4$.  It
can also be shown that if a manifold has precisely one covariantly
constant spinor $\epsilon$, then its holonomy group is $G_2$, or at
least a large subgroup thereof.  One can in fact construct the
covariant three-form $\varphi$ as a bilinear in $\epsilon$.

It would seem that $G_2$ holonomy compactifications of 11-dimensional
supergravity would be of utmost phenomenological interest; however,
one should recall Witten's proof \cite{WittenShelter} that
compactifications of 11-dimensional supergravity on any {\it smooth}
seven-manifold cannot lead to chiral matter in four dimensions.  With
a modern perspective, we conclude that we should therefore be studying
{\it singularities} in $G_2$ holonomy manifolds, or branes, or some
other defects where chiral fermions might live.

\begin{figure}[h]
  \centerline{\psfig{figure=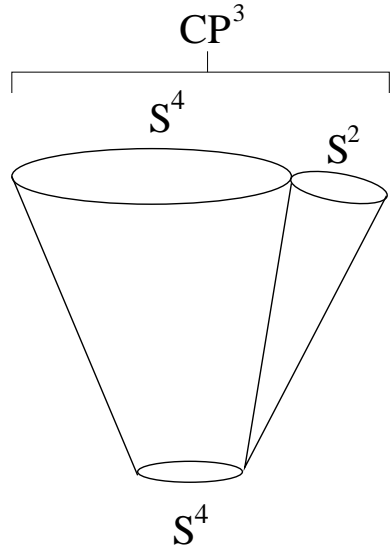,width=2in}}
   \caption{${\bf CP}^3$ a $S^2$ fibration over $S^4$, and in the
interior, the $S^2$ shrinks but the $S^4$ doesn't.}
   \label{figI}
  \end{figure}

The usual starting point for investigating singularities in an
$n$-dimensional manifold is to look at non-compact manifolds which are
asymptotically conical: 
  \eqn{ConeSing}{
   ds_n^2 \sim dr^2 + r^2 d\Omega_{n-1}^2
  }
 for large $r$.  Note that if $\sim$ were replaced by an exact
equality, then the metric $ds_n^2$ would be singular at $r=0$ unless
$d\Omega_{n-1}^2$ is the metric on a unit $(n-1)$-sphere.  In the
previous lecture, we encountered a prime example of this sort of
singularity: $A_k$ singularities in four-manifolds locally have the
form \ConeSing\ with $d\Omega_3^2$ being the metric of the Lens space
$S^3/{\bf Z}_{k+1}$.  Another frequently discussed example is the
conifold singularity in Calabi-Yau three-folds: this is locally a cone
over the coset space $T^{11} = SU(2) \times SU(2) / U(1)_{\rm diag}$.
The conifold admits a Calabi-Yau metric that is known explicitly, as
are certain resolutions of the singularity which remain Calabi-Yau
(much like the resolutions of the $A_k$ singularities discussed in the
previous lecture).  As remarked previously, one can gain tremendous
insight into Calabi-Yau singularities through algebraic equations: for
instance, the $A_1$ space and the conifold can be described,
respectively, via the equations $\sum_{i=1}^3 z_i^2 = 0$ and
$\sum_{i=1}^4 z_i^2 = 0$.  Sadly, there is no such algebraic tool
known for describing singular or nearly singular $G_2$ holonomy
manifolds.  And in fact, there are essentially only three known
asymptotically conical metrics of $G_2$ holonomy.  The bases of the
cones are ${\bf CP}^3$, ${SU(3) \over U(1) \times U(1)}$, and $S^3
\times S^3$, but the metrics $d\Omega_6^2$ that appear through
\ConeSing\ in the $G_2$ holonomy metrics are not the obvious metrics
on these spaces (just as, in fact, the metric on $T^{11}$ induced by
the Calabi-Yau metric on the conifold is not quite the metric
suggested by the coset structure).  The three metrics admit isometry
groups $SO(5)$, $SU(3)$, and $SU(2)^3$, respectively.  (Don't get
confused between isometry and holonomy: isometry means that after some
transformation the metric is the same as before, whereas holonomy
tells us how complicated the transformation properties of vectors are
under parallel transport).  And at the ``tip'' of the three respective
asymptotically conical metrics, an $S^4$, or a ${\bf CP}^2$, or an
$S^3$, remains finite.  See figure~\ref{figI} for a schematic
depiction of the $S^4$ case.

We may describe the explicit $G_2$ holonomy metrics in terms only
slightly more complicated than the explicit metric \EHTwo\ for
$EH_2$.  For the $SO(5)$ symmetric case, one has
  \eqn{CPThreeCase}{
   ds_7^2 = {dr^2 \over 1 - r^{-4}} + 
    {1 \over 4} r^2 (1 - r^{-4}) (d\mu^i + \epsilon^{ijk} A^j \mu^k)^2 + 
    {1 \over 2} r^2 ds_4^2 \,,
  }
 where $ds_4^2$ is the $SO(5)$ symmetric metric on a unit $S^4$, the
$\mu^i$ are three Cartesian coordinates on $S^2$, subject to
$\sum_{i=1}^3 (\mu^i)^2 = 1$, and $A^i_\mu$ is an $SU(2)$ gauge field
on $S^4$ carrying unit instanton number.  We can be a little more
explicit about this gauge field, as follows.  $S^4$ is a space of
$SO(4)$ holonomy, but $SO(4) \approx SU(2)_L \times SU(2)_R$, and the
spin connection $\omega_{\mu\ ab}$ is decomposable into $SU(2)_L$ and
$SU(2)_R$ pieces as $\omega_{\mu\ ab}^{L,R} = \omega_{\mu\ cd} \left(
\delta^c_a \delta^d_b \pm {1 \over 2} \epsilon^{cd}{}_{ab} \right)$.
The gauge field $A^i_\mu$ can be taken proportional to $\sigma^i_{ab}
\omega_{\mu\ ab}^L$, where $\sigma^i_{ab}$ are the Pauli matrices.
The ${\bf CP}^2$ case is identical to the above discussion, only one
takes $ds_4^2$ to be the $SU(3)$ symmetric metric on a ${\bf CP}^2$
whose size is such that the Ricci curvature is three times the metric
(as is true for a unit $S^4$).  Clearly, when $r=1$, the $S^2$ part of
the metric shrinks to nothing, while the $S^4$ or ${\bf CP}^2$ remains
finite.  Topologically, the whole space is a bundling of ${\bf R}^3$
over $S^4$ or ${\bf CP}^2$, and the base is the corresponding $S^2$
bundle over $S^4$ or ${\bf CP}^2$.

The $SU(2)^3$ symmetric case is actually more ``elementary,'' in the
sense that we do not need to discuss gauge fields.  The metric is
  \eqn{SThreeCase}{
   ds_7^2 = {dr^2 \over 1 - r^{-3}} + {1 \over 9} r^2 (1-r^{-3})
    (\nu_1^2 + \nu_2^2 + \nu_3^2) + 
    {r^2 \over 12} (\sigma_1^2 + \sigma_2^2 + \sigma_3^2)
  }
 where $\nu_i = \Sigma_i - {1 \over 2} \sigma_i$, and $\Sigma_i$ and
$\sigma_i$ are left-invariant one-forms on two different $S^3$'s.
Clearly, only one of these $S^3$'s stays finite as $r \to 1$.
Topologically, the whole space is a bundling of ${\bf R}^4$ over
$S^3$.  Any $G_2$ holonomy metric can be rigidly rescaled without
changing its holonomy group: thus we can claim to have exhibited three
one-parameter families of $G_2$ holonomy metrics, each parametrized by
the $S^4$, or ${\bf CP}^2$, or $S^3$, that remains unshrunk.  A
perfectly conical metric has an isometry under scaling: $dr^2 + r^2
d\Omega_6^2 \to d(\Omega r)^2 + (\Omega r)^2 d\Omega_6^2$ for any
constant factor $\Omega$.  Thus the asymptotics of the rescaled
metrics is always the same.  And the limit where the unshrunk space at
the center goes to zero volume is an exactly conical metric.
Considerably more detail on these $G_2$ holonomy metrics can be found
in the original papers \cite{BryantSal,GPP}.

It may seem that our discussion of $G_2$ holonomy metrics is
remarkably unenlightening and difficult to generalize.  This is true!
Despite more than 15 years since the discovery of the metrics
\CPThreeCase\ and \SThreeCase, there are few generalizations of them,
and little else known about explicit $G_2$ holonomy metrics.  One
interesting generalization of \SThreeCase\ is the discovery of less
symmetric versions where, as with Taub-NUT space, there is a $U(1)$
fiber which remains finite at infinity \cite{BGGG}.  Nevertheless,
there are several generally useful observations to make at this point:
  \begin{itemize}
 \item $G_2$ holonomy implies Ricci flatness.  A mathematically
rigorous proof is straightforward, but a nice physical argument is
that $G_2$ holonomy on $M_7$ implies unbroken supersymmetry for
eleven-dimensional supergravity on ${\bf R}^{3,1} \times M_7$ with
$G_{(4)}=0$; and supersymmetry implies the equations of motion, which
for $G_{(4)}=0$ are precisely Ricci flatness.  Ricci flatness is a common
feature of special holonomy manifolds: $SU(n)$ and $Spin(7)$ holonomy
manifolds are also necessarily Ricci-flat; but $U(n)$ holonomy
manifolds are not.
 \item The condition $\nabla_\mu \varphi_{\alpha\beta\gamma}=0$ can be
shown to be {\it equivalent} to the apparently {\it weaker} condition
$d\varphi = 0 = d*\varphi$.  These first order equations can be
considerably easier to solve than $R_{\mu\nu}=0$.  The three-forms for
each of the three ``classical'' $G_2$ holonomy metrics are explicitly
known, but we would not gain much from exhibiting their explicit
forms.
 \item The three-form $\varphi$, as well as its Hodge dual $*\varphi$,
are calibrations.  Examples of calibrated three- and four-cycles are
the unshrunk $S^3$, $S^4$, and ${\bf CP}^2$ at $r=1$ in the metrics
\CPThreeCase\ and \SThreeCase.  An M2-brane on the unshrunk $S^3$
would be a supersymmetric instanton in M-theory, similar to the
worldsheet instantons arising from strings on holomorphic curves.  An
exploration of such instantons (including their zero modes) can be
found in \cite{HarveyMoore}.
 \item M-theory has a 3-form potential, $C$.  Just as we formed $J+iB$
in string theory, so we can form $\varphi+i C$, and then $\int_{S^3}
(\varphi+iC)$ is the analog of a complexified Kahler parameter
$\int_{N^A} (J+iB)$.  As an example of the use of this analogy, one
may show that $M2-brane$ instantons make a contribution to the
superpotential whose dominant behavior is $\exp\left\{ -\tau_{M2}
\int_{S^3} (\varphi+iC) \right\}$.  Perturbative corrections to the
classical superpotential are forbidden by the usual holomorphy plus
Peccei-Quinn symmetry argument.  However, in contrast to the case of
type II superstrings, or heterotic strings with the standard
embedding, where the Kahler potential could be related to a
holomorphic prepotential, it is difficult to say anything systematic
about the Kahler potential for $G_2$ compactifications: no ``hidden''
supersymmetry is available, and perturbative corrections at all orders
seem to be allowed.
 \end{itemize}

Recall that after discussing the $A_1$ singularity in detail, we were
able to go on to construct a smooth, compact $SU(2)$ holonomy manifold
by resolving $A_1$ singularities of an orbifold of $T^4$ by a discrete
subgroup of $SU(2)$.  Around each fixed point, we cut out little
regions of the orbifold, whose local geometry was $B^4/{\bf Z}_2$
($B^4$ being a unit ball), and we replaced them by cut-off copies of
the Eguchi-Hansen space $EH_2$.  Smoothing out the small
discontinuities in the metric at the joining points, without losing
$SU(2)$ holonomy, was an interesting subtlety that we left for the
mathematical literature.  It turns out that a very similar strategy
suffices to construct smooth compact $G_2$ holonomy manifolds.  This
is called the Joyce construction, and it was the way in which the
first explicit examples of compact $G_2$ holonomy manifolds were found
\cite{JoyceOne,JoyceTwo}.  We start with an orbifold $T^7/\Gamma$,
where $T^7$ is the square unit torus parametrized by $\vec{x} =
(x^1,\ldots,x^7)$, and $\Gamma$ is a discrete subgroup of $G_2$, to be
specified below.  $\Gamma$ has a set of fixed points $S$ which, in the
upstairs picture, is locally a three-dimensional submanifold of $T^7$.
Each fixed point is an $A_1$ singularity.  The key step is to replace
$S \times B^4/{\bf Z}_2$ by $S \times EH_2$, and then argue that after
smoothing out the small discontinuities, the resulting smooth manifold
has $G_2$ holonomy.

A particular example of this strategy begins with the discrete
subgroup $\Gamma$ generated by the following three transformations:
  \eqn{abc}{\eqalign{
   \alpha:&\ \vec{x} \to (-x^1,-x^2,-x^3,-x^4,x^5,x^6,x^7)  \cr
   \beta:&\ \vec{x} \to (-x^1,\textstyle{1 \over 2} - x^2,x^3,x^4,
     -x^5,-x^6,x^7)  \cr
   \gamma:&\ \vec{x} \to (\textstyle{1 \over 2} - x^1,x^2,
     \textstyle{1 \over 2} - x^3,x^4,-x^5,x^6,-x^7) \,.
  }}
 These generators have several nice properties which make the Joyce
construction work:
  \begin{itemize}
 \item They commute.  The group $\Gamma$ is ${\bf Z}_2^3$.
 \item They preserve a three-form $\varphi$ of the form \PhiLocal\
(with an appropriate relabellings of the $x^i$'s as $y^j$'s), so
indeed the action of $\Gamma$ induced by \abc\ on the tangent space of
$T^7$ is a subgroup of the usual action of $G_2 \subset SO(7)$.  (This
is what we mean, precisely, by $\Gamma \subset G_2$).
 \item The generators $\alpha$, $\beta$, and $\gamma$ each
individually has a fixed point set in $T^7$ consisting of 16 $T^3$'s.
$\beta$ and $\gamma$ act freely on the fixed point set of $\alpha$,
and similarly for the fixed point sets of $\beta$ and $\gamma$.
 \item The 48 $T^3$'s coming from the fixed point sets of the
generators $\alpha$, $\beta$, and $\gamma$ are disjoint, but the 16
from $\alpha$ are permuted by $\beta$ and $\gamma$, and similarly for
the 16 from $\beta$ and from $\gamma$.  Thus on the quotient space,
$S$ consists of 12 disjoint $T^3$'s.
  \end{itemize}

Since $S$ has 12 disjoint components in the quotient space, we must
ensure when replacing $S \times B^4/{\bf Z}_2$ by $S \times EH_2$ that
all 12 $SU(2)$'s are in the {\it same} $G_2$.  To this end, we exploit
the fact that $EH_2$ is hyperkahler, which is to say its metric is
Kahler with respect to three different complex structures.  In
practice, what this means is that there exist covariantly constant
$\omega_1$, $\omega_2$, and $\omega_3$ (the three possible Kahler
forms), which in local coordinates at any given point can be written
as
  \eqn{ThreeOmegas}{\eqalign{
   \omega_1 &= dy^1 \wedge dy^4 + dy^2 \wedge dy^3  \cr
   \omega_2 &= dy^1 \wedge dy^3 - dy^2 \wedge dy^4  \cr
   \omega_3 &= dy^1 \wedge dy^2 + dy^3 \wedge dy^4 \,.
  }}
 (On the singular space ${\bf C}^2/{\bf Z}_2$, the $y^i$ could be
taken as real coordinates for ${\bf C}^2$).  Now, the cotangent space
of any one of the 12 $T^3$'s is spanned by three one-forms: $dx^i$,
$dx^j$, and $dx^k$ for some choice of $i$, $j$, and $k$.  For a
correct ordering of $i$, $j$, and $k$, and correct identification of
the $y$ coordinates in \ThreeOmegas\ with the remaining four $x$
coordinates, the form
  \eqn{PhiAgain}{
   \varphi = \omega_1 \wedge dx^1 + \omega_2 \wedge dx^2 + 
    \omega_3 \wedge dx^3 + dx^1 \wedge dx^2 \wedge dx^3
  }
 is precisely the original three-form \PhiLocal, written at the
location of each $T^3$ in a way which the replacement $B^4/{\bf Z}_2
\to EH_2$ clearly preserves.  This is the reasoning that allows us to
say that the holonomy group is still contained in $G_2$ after the
resolution.  As before, we gloss over the subtlety of smoothing out
the discontinuities; this is well treated in Joyce's original papers
\cite{JoyceOne,JoyceTwo}.  There it is also shown that the moduli
space of $G_2$ metrics is locally $H^3(M_7,{\bf R}) = {\bf R}^{43}$
for this example.  The moduli space of M-theory on this manifold is
locally $H^3(M_7,{\bf C})$ because of the complexification
$\varphi+iC$.  The Kahler potential on this moduli space is probably
{\it hard} to compute beyond the classical level, for the reasons
explained above.

Beautiful and impressive though the Joyce construction is, we still
seem as yet rather stuck in mathematics land in our study of $G_2$
holonomy manifolds.  There are two main themes in the relation of
$G_2$ holonomy to string theory.  One, which we will not discuss,
centers on a relationship with strongly coupled gauge theories,
developed in \cite{Vafa,Acharya,WittenAtiyah}.  The other, perhaps
more obvious relation, is with configurations of D6-branes in type IIA
string theory.  To begin, we should recall the basic ansatz relating
type IIA string theory to M-theory:
  \eqn{MetricAnsatz}{
   ds_{11}^2 = e^{-2\Phi/3} ds_{str}^2 + e^{4\Phi/3} (dx^{11} + 
    C_\nu dx^\nu)^2 \,,
  }
 where $C_\mu$ is the Ramond-Ramond one-form of type IIA, and $\Phi$
is the dilaton.  The classical geometry for $n+1$ flat, parallel
D6-branes can be cast in the form \MetricAnsatz:
  \eqn{TNRSix}{\eqalign{
   & ds_{11}^2 = ds_{{\bf R}^{6,1}}^2 + H d\vec{r}^{\,2} + 
    H^{-1} (dx^{11} + \vec\omega \cdot d\vec{r})^2  \cr
   &\quad \nabla \times \vec\omega = -\nabla H \qquad
    e^\Phi = H^{-3/4} \qquad 
    H = 1 + {1 \over 2} \sum_{i=1}^{n+1} 
     {R \over |\vec{r}-\vec{r}_i|}  \cr
   & ds_{str}^2 = H^{-1/2} ds_{{\bf R}^{6,1}}^2 + H^{1/2} d\vec{r}^{\,2} 
     \qquad R = g_{str} \sqrt{\alpha'} \,.
  }}
 Here $\vec{r}$ parametrizes the three directions perpendicular to the
D6-branes, whose centers are at the various $\vec{r}_i$.  Since the
eleven-dimensional geometry is the direct product of flat ${\bf
R}^{6,1}$ and multi-center Taub-NUT, the holonomy group is $SU(2)$,
and hence $1/2$ of supersymmetry is preserved.  It is indeed
appropriate, since parallel D6-branes preserve this much
supersymmetry.  A more general observation is that, since D6-branes
act as sources only for the metric, the Ramond-Ramond one-form, and
the dilaton, and these fields are organized precisely into the
eleven-dimensional metric, {\it any} configuration of D6-branes that
solves the equations of motion must lift to a Ricci-flat manifold in
eleven dimensions; and if the configuration of D6-branes is
supersymmetric, then the eleven-dimensional geometry must have at
least one covariantly constant spinor, and hence special holonomy.  In
particular, if there is a factor of flat ${\bf R}^{3,1}$ in the
geometry, and some supersymmetry is unbroken the rest of it must be a
seven-manifold whose holonomy is contained in $G_2$.  In the example
above, the seven-manifold is ${\bf R}^3$ times multi-center Taub-NUT.

Before developing this theme further, it seems worthwhile to explore
the dynamics of $n+1$ parallel D6-branes, as described in \TNRSix, a
little further.  Recall that we learned in lecture~1 that there are
$n$ homologically non-trivial cycles for the $n+1$-center Taub-NUT
geometry: topologically, this is identical to a resolved $A_n$
singularity.  Thus there exist $n$ harmonic, normalizable two-forms,
call them $\omega^i$.  These forms are localized near the centers of
the Taub-NUT space, and they are the cohomological forms dual to the
$n$ non-trivial homology cycles.  Furthermore, there is one additional
normalizable 2-form on the Taub-NUT geometry, which can be constructed
explicitly for $n=0$, but which owes its existence to no particular
topological property.  Let us call this form $\omega^0$.  If we expand
the Ramond-Ramond three-form of type IIA as
  \eqn{CExpand}{
   C_{(3)} = \sum_{i=0}^n \omega^i \wedge A_i + \ldots \,,
  }
 where the $A_i$ depend only on the coordinates of ${\bf R}^{6,1}$,
then each term represents a seven-dimensional $U(1)$ gauge field
localized on a center of the Taub-NUT space.  This is very
appropriate, because there is indeed a $U(1)$ gauge field on each
D6-brane: through \CExpand\ we are reproducing this known fact from
M-theory.  Better yet, recall that there are $n(n+1)/2$ holomorphic
embeddings of $S^2$ in a $n+1$-center Taub-NUT space.  An M2-brane
wrapped on any of these is some BPS particle, and an anti-holomorphic
wrapping is its anti-particle.  A closer examination of quantum
numbers shows that these wrapped M2-branes carry the right charges and
spins to be the non-abelian $W$-bosons that we know should exist: in
the type IIA picture they are the lowest energy modes of strings
stretched from one D6-brane to another.  It's easy to understand the
charge quantum number for the case $n=1$, that is for two D6-branes.
The form $\omega^0$ corresponds to what we would call the
center-of-mass $U(1)$ of the D6-branes.  The form $\omega^1$ is dual
to the holomorphic cycle over which we wrap an M2-brane: this is
precisely the holomorphic $S^2$ at $r=\epsilon$ in the Eguchi-Hansen
space \EHTwo, as discussed after \JIneq.  Thus $\int_{S^2} \omega^1 =
1$, which means that an M2-brane on this $S^2$ does indeed have charge
$+1$ under the $U(1)$ photon which we called $A_1$ in \CExpand.  This
is the relative $U(1)$ in the D6-brane description, and the wrapped
M2-brane becomes a string stretched between the two D6-branes, which
does indeed have charge under the relative $U(1)$.

\begin{figure}[h]
   \vskip-0.8in
   \centerline{\psfig{figure=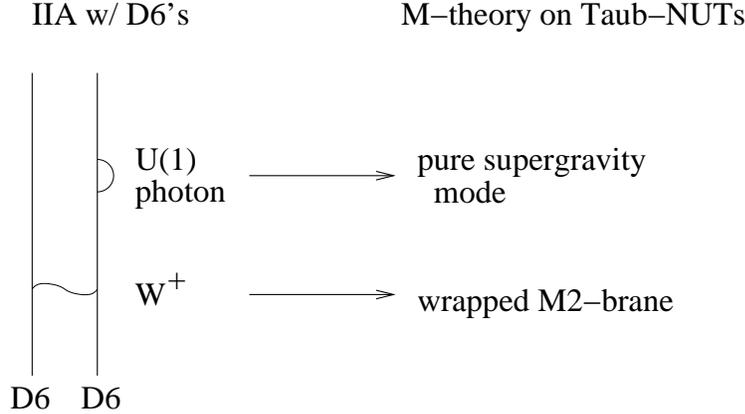,width=5in,angle=270}}
   \vskip-0.8in
   \caption{Stretched string degrees of freedom in type IIA lift to
very different things in M-theory: a $U(1)$ photon with both its ends
on a single D6-brane lifts to a zero-mode of $C_{MNP}$, whereas a
$W^+$, with one end on one D-brane and the other on another, lifts to
a wrapped M2-brane.  These states can be tracked reliably from weak to
strong string coupling because they are BPS.}
   \label{figG}
  \end{figure}

When two D6-branes come together, one of the holomorphic cycles
shrinks to zero size, and there is gauge symmetry enhancement from
$U(1) \times U(1)$ to $U(2)$.  In the generic situation where the
D6-branes are separated, the unbroken gauge group is $U(1)^{n+1}$ on
account of the Higgs mechanism.  This is a pretty standard aspect of
the lore on the relation between M-theory and type IIA, but I find it
a particularly satisfying result, because it shows that wrapped
M2-branes have to be considered on an equal footing with the degrees
of freedom of eleven-dimensional supergravity: in this instance, they
conspire to generate $U(n)$ gauge dynamics.  See figure~\ref{figG}.
 Such a conspiracy is one of the reasons why we believe the type
duality between IIA string theory and 11-dimensional M-theory goes
beyond the supergravity approximation.

Let's return to our earlier observation that any configuration of
D6-branes in type IIA must lift to pure geometry in eleven dimensions.
Actually, more is true: any type IIA configuration that involves only
the metric, the dilaton, and the Ramond-Ramond one-form should lift to
pure geometry in eleven dimensions.  This means we can include
O6-planes as well as D6-branes.  Our focus here, however, will be on
D6-branes only.  Consider, in particular, a set of D6-branes which all
share a common ${\bf R}^{3,1}$, which we could regard as our own four
dimensions.  Assume that the configuration preserves ${\cal N}=1$
supersymmetry in $d=4$.  Then the lift to eleven dimensions should
generically be ${\bf R}^{3,1}$ times a $G_2$ holonomy manifold.  (We
have not entirely ruled out cases where the holonomy group is smaller
than $G_2$, but we expect such configurations to be quite special, if
they exist at all).

Suppose, for instance, that the other six dimensions in the type IIA
description are non-compact and asymptotically flat (or else
compact/curved on a much larger length scale than we are considering
for now), and that each D6-brane stretches along some flat ${\bf
R}^3 \subset {\bf R}^6$.  When would this configuration preserve
${\cal N}=1$ supersymmetry?  The answer to this sort of question was
given in one of the early papers on D-branes
\cite{bdl}, and it relies on a fermionic Fock-space
trick which is also of use in the study of spinors and differential
forms on Calabi-Yau spaces.  The best way to state the result of
\cite{bdl} is to first choose complex coordinates on
${\bf R}^6$, call them $z^1$, $z^2$, and $z^3$.  Obviously there are
many inequivalent ways to form the $z^i$, but suppose we've made up
our mind on one for the moment.  Now, the $SU(3)$ acting on the $z^i$
by
  \eqn{RAction}{
   z^i \to R^i{}_j z^j \qquad 
   \bar{z}^{\bar{i}} \to R^{\dagger\,\bar{i}}{}_{\bar{j}} \bar{z}^{\bar{j}} \,,
  }
 is obviously a subgroup of all possible $SO(6)$ rotations.  Let us
construct Dirac gamma matrices obeying
$\{\Gamma^{z^i},\Gamma^{\bar{z}^{\bar{j}}}\} = 2 g^{i\bar{j}} \propto
\delta^{i\bar{j}}$.  Clearly, the $\Gamma^{z^i}$ and
$\Gamma^{\bar{z}^{\bar{j}}}$ are fermionic lowering/raising operators,
up to a normalization.  We can define a Fock space vacuum $\epsilon$
in the spinor representation of the Clifford algebra via $\Gamma^{z^i}
\epsilon = 0$ for all $i$.  The full Dirac spinor representation of
$SO(6)$ now decomposes under the inclusion $SU(3) \subset SO(6)$ as
${\bf 8} = {\bf 1} \oplus \bar{\bf 3} \oplus {\bf 3} \oplus {\bf 1}$.
The singlets are $\epsilon$ itself and $\Gamma^{\bar{z}^1}
\Gamma^{\bar{z}^2} \Gamma^{\bar{z}^3} \epsilon$; the $\bar{\bf 3}$ is
$\Gamma^{\bar{z}^{\bar{j}}} \epsilon$; and the ${\bf 3}$ is
$\Gamma^{\bar{z}^{\bar{i}}} \Gamma^{\bar{z}^{\bar{j}}} \epsilon$.

With this mechanism in place, we can state and immediately understand
the results of \cite{bdl}: suppose one D6-brane lies
along the ${\bf R}^3$ spanned by the real parts of $z^1$, $z^2$, and
$z^3$.  Consider a collection of other D6-branes on ${\bf R}^3$'s
related by various $SU(3)$ rotations, and, optionally, arbitrary
translations.  The claim is that this configuration preserves ${\cal
N}=1$ supersymmetry.  To understand why this is so, we need only
recall that the first D6-brane preserves the half of supersymmetry
satisfying
  \eqn{FirstDSix}{
   \tilde{\epsilon}_R = \prod_{i=1}^3 \left( \Gamma^{z^i} + 
    \Gamma^{\bar{z}^{\bar{i}}} \right) \epsilon_L \,,
  }
 where $\tilde{\epsilon}_R$ is the right-handed spacetime spinor that
comes from the anti-holomorphic sector on the worldsheet, and
$\epsilon_L$ is the left-handed spacetime spinor that comes from the
holomorphic sector.  The rotated D6-branes also preserve half of
supersymmetry, but a different half, namely
  \eqn{OtherDSix}{
   \tilde{\epsilon}_R = \prod_{i=1}^3 \left( R^i{}_j \Gamma^{z^j} + 
    R^{\dagger\,\bar{i}}{}_{\bar{j}} \Gamma^{\bar{z}^{\bar{j}}} \right) 
    \epsilon_L \,.
  }
 Some supersymmetry is preserved by the total collection of D6-branes
iff we can find simultaneous solutions to \FirstDSix\ and \OtherDSix\
for the various $SU(3)$ rotations $R^i{}_j$ that appear.  In fact, if
$\epsilon_L$ is the Fock space vacuum $\epsilon$ tensored with an
arbitrary chiral spinor in four-dimensions, and $\tilde{\epsilon}_R$
is $\Gamma^{\bar{z}^1} \Gamma^{\bar{z}^2} \Gamma^{\bar{z}^3} \epsilon$
times the same four-dimensional chiral spinor, then \FirstDSix\ is
obviously satisfied; but also \OtherDSix\ is satisfied, because
  \eqn{GetTilde}{
   \tilde{\epsilon}_R = \prod_{i=1}^3 R^{\dagger\,\bar{i}}{}_{\bar{j}}
    \Gamma^{\bar{z}^{\bar{j}}} \epsilon_L = 
    (\det R^{\dagger\,\bar{i}}{}_{\bar{j}}) 
     \Gamma^{\bar{z}^1} \Gamma^{\bar{z}^2} \Gamma^{\bar{z}^3} 
     \epsilon_L
  }
 when $\epsilon_L$ is as stated above; and $\det R^{\dagger\,
\bar{i}}{}_{\bar{j}} = 1$ for an $SU(3)$ matrix.

\begin{figure}[h]
   \centerline{\psfig{figure=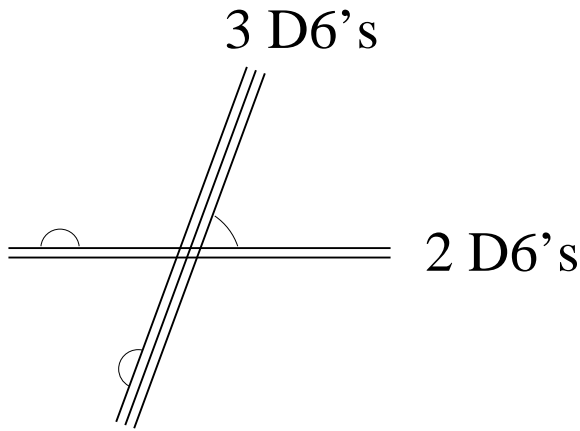,width=3in}}
   \caption{Three coincident D6-branes intersecting two coincident
D6-branes at a unitary angle.  Strings are shown that give rise to
$U(3) \times U(2)$ gauge fields (on the respective D6-brane
worldvolumes) and chiral fermions in the $({\bf 3},\bar{\bf 2})$ (at
the intersection).}
   \label{figH}
  \end{figure}

More in fact was shown in \cite{bdl}: it turns out that ${\cal N}=1$
$d=4$ chiral matter lives at the intersection of D6-branes oriented at
unitary angles in the manner discussed in the previous paragraph.  We
will not here enter into the discussion in detail, but merely state
that the GSO projection that acts on string running from one D6-brane
to another one at a unitary angle from the first winds up projecting
out the massless fermions of one chirality and leaving the other.
Clearly, such strings carry bifundamental charges under the gauge
group on the D6-branes they end on; so if one intersects, say, a stack
of two coincident D6-branes with another stack of three, the
four-dimensional dynamics of the intersection is $U(3) \times U(2)$
gauge theory with chiral ``quarks'' in the $({\bf 3},\bar{\bf 2})$.
This is remarkably similar to the Standard Model!  See
figure~\ref{figH}.
 After these lectures were given, work has appeared \cite{Cvetic}
where explicit compact constructions are given, involving D6-branes
and O6-planes, whose low-energy dynamics includes the supersymmetric
Standard Model (as well as some other, possibly innocuous, extra
degrees of freedom).  See also the related work \cite{WittenAcharya}.
As is clear from the previous discussion, such constructions lift in
M-theory to configurations which involve only the metric, not
$G_{(4)}$.  Only there are singularities in the eleven-dimensional
metric where D6-branes cross.  As shown in \cite{WittenAtiyah}, the
singularity in the $G_2$ holonomy lift of two D6-branes intersecting
at unitary angles is precisely the cone over ${\bf CP}^3$ exhibited in
\CPThreeCase\ (or, properly, the limit of this geometry where the
$S^4$ shrinks); whereas the singularity in the $G_2$ holonomy lift of
of three D6-branes (all at different, unitarily related angles) is the
cone over ${SU(3) \over U(1) \times U(1)}$ (of a form very similar to
\CPThreeCase, as discussed above).  More detail can be found in
\cite{WittenAtiyah}, and also in the earlier work
\cite{KachruMcGreevy}, on the D6-brane interpretation of resolving the
conical singularity.  Oddly, it seems that the generalizations of
these $G_2$ holonomy cones to any number of D6-branes, intersecting
all at different angles, is not known.  Also, the full geometry
interpolating between the near-intersection region, where the metric
is nearly conical, and the asymptotic region, where the metric is
nearly Taub-NUT close to any single D6-brane, is not known.  Finding
either sort of generalization of the existing result \CPThreeCase\
would be very interesting, and possibly useful for studying the
dynamics of M-theory compactifications.

A word of explanation is perhaps in order on why we have focused so
exclusively on M-theory geometries with $G_{(4)} = 0$.  Really there
are two.  First, on a compact seven-manifold, there are rather tight
constraints on how $G_{(4)}$ may be turned on---see for example
\cite{BeckerSeven}.  The main context of interest where non-zero
$G_{(4)}$ seems necessary is compactifications of Horava-Witten
theory: there is seems impossible to satisfy the anomaly condition
$\tr R \wedge R - {1 \over 2} \tr F \wedge F = 0$ on each $E_8$ plane
individually, so some $G_{(4)}$ is needed to ``soak out'' the anomaly.
The second reason to consider M-theory geometries first with $G_{(4)}
= 0$ is that, in an expansion in the gravitational coupling, the
zeroth order equations of motion are indeed Ricci-flatness.  For
instance, in Horava-Witten compactifications, the necessary $G_{(4)}$
is only a finite number of Dirac units through given four-cycles.  As
long as only finitely many quanta of $G_{(4)}$ are turned on, and
provided the compactification scale is well below the
eleven-dimensional Planck scale, one would expect to learn much by
starting out ignoring $G_{(4)}$ altogether.  This is not to say that
nonzero $G_{(4)}$ won't have some interesting and novel effects: see
for example \cite{beckerEight,cglp}.  It is fair to say that our
understanding of M-theory compactifications is in a very primitive
state, as compared, for instance, to compactifications of type~II or
heterotic strings.  It is to be hoped that this topic will flourish in
years to come.

\subsection{Addendum: further remarks on intersecting D6-branes}
\label{DSix}

My original TASI lectures ended here, but in view of the continuing
interest in $G_2$ compactifications of M-theory, it seems worthwhile
to present a little more detail on intersecting D6-branes and their
M-theory lift.  This in fact is a subject where rather little is
known, so I will in part be speculating about what might be
accomplished in further work.

First it's worthwhile to reconsider the work of \cite{bdl} in light of
a particular calibration on ${\bf R}^6$.  Consider the complex
three-form $\Omega = dz^1 \wedge dz^2 \wedge dz^3$, where the $z^i$
are, as before, complex coordinates on ${\bf R}^6$ such that the
metric takes the standard Kahler form.  $\Omega$ is called the
holomorphic three-form, or the holomorphic volume form, and if space
had permitted, some elegant results could have been presented about
how the analogous form on a curved complex manifold relates to its
complex structure, as well as to covariantly constant spinors, if they
exist.  Our purpose here is to note that $\Re\Omega$ is a calibration,
in the sense explained in section~\ref{Sigma}.  Clearly $\Re\Omega$
calibrates the plane in the $\Re z^1$, $\Re z^2$, $\Re z^3$
directions.  Any $SU(3)$ change of the coordinates $z^i$ preserves
$\Omega$; in fact such a map is the most general linear map that does
so.  So it is not hard to convince oneself that all planes related to
the one we first mentioned by a $SU(3)$ rotation are also calibrated
by $\Re\Omega$.  One can now concisely restate the result that
D6-branes stretched on ${\bf R}^{3,1}$ must be at unitary angles in
the remaining ${\bf R}^6$ to preserve supersymmetry: supersymmetric
intersecting D6-branes must all be calibrated by $\Re\Omega$, for some
suitable choice of the $z^i$.  Choice of the $z^i$ here includes the
ability to rotate $\Re\Omega$ by a phase.  A more general truth is
that supersymmetric D6-branes on a three-cycle of a Calabi-Yau
manifold are those calibrated by $\Re\Omega$.  Such three-cycles are
called special lagrangian manifolds.\footnote{We have not been quite
precise in the main text: more accurately, D6-branes should wrap
special lagrangian manifolds, and such a manifold has the properties
that both the Kahler form and $\Im\Omega$ pull back to zero, as well
as being calibrated by $\Re\Omega$.}

It can be shown (see for example \cite{WittenAtiyah}) that the $G_2$
cones over ${\bf CP}^3$ and ${SU(3) \over U(1) \times U(1)}$ are
limits of the M-theory lift of two and three D6-branes intersecting at
a common point and at unitary angles.  It is even known how the $G_2$
resolutions of these cones corresponds to deformed world-volumes of
the D6-branes: for instance, for the cone over ${\bf CP}^3$, the
D6-brane world-volume has an hour-glass shape which is topologically
$S^2 \times {\bf R}^2$.

We can describe in a straightforward fashion, though without
mathematical rigor, how supersymmetric D6-brane configurations
spanning ${\bf R}^{3,1}$ and a special lagrangian manifold in ${\bf
R}^6$ can be lifted to manifolds of $G_2$ holonomy in eleven
dimensions.  As explained in section~\ref{GTwo}, the
eleven-dimensional lift of a single flat D6-brane is Taub-NUT space
times ${\bf R}^{6,1}$, which we will conveniently write as ${\bf
R}^{3,1} \times {\bf R}^3$.  Intuitively speaking, we should be able
to lift any D6-brane configuration with no coincident or intersecting
D6-branes, just by making an affine approximation to the curving
D6-brane world volume at each point, and lifting to Taub-NUT times the
D6-brane world-volume times ${\bf R}^{3,1}$.  From now own let's
ignore the ${\bf R}^{3,1}$ part.  Then in the seven remaining
dimensions, the geometry far from any brane is ${\bf R}^6 \times S^1$.
Near the D6-brane world volume, we cut out a region in ${\bf R}^6$
that surrounds the brane, and since locally this region is ${\bf R}^3
\times B^3$, we can replace ${\bf R}^3 \times B^3 \times S^1$ in the
seven dimensional geometry by ${\bf R}^3$ times a cut-off Taub-NUT.
Gluing in the Taub-NUT space should cause only very small
discontinuities, which hopefully could be erased through some real
analysis.

There is a meaningful point to check, though: in our putative
almost-$G_2$ manifold, formed by gluing into ${\bf R}^6 \times S^1$ a
cut-off Taub-NUT snaking along what was the D6-brane world volume,
we'd like to see that the holonomy on different parts of the ``snake''
is always (nearly) contained in the same $G_2$.  To this end, we need
to write down a covariantly constant three-form $\varphi$ for ${\bf
R}^3$ times Taub-NUT.  This can be done in different ways, because
there are different embeddings of $SU(2)$ into $G_2$.  Let $x^1$,
$x^2$, and $x^3$ be coordinates on ${\bf R}^3$, and let $y^1$, $y^2$,
$y^3$, and $x^{11}$ be coordinates on Taub-NUT.  One choice suggested
by the discussion in section~\ref{GTwo} is to use the fact that
Taub-NUT is hyperkahler, and construct
  \eqn{OnePhi}{
   \varphi = \omega_1 \wedge dx^1 + \omega_2 \wedge dx^2 + 
    \omega_3 \wedge dx^3 + dx^1 \wedge dx^2 \wedge dx^3 \,,
  }
 where the $\omega_i$ are the Kahler structures.  This is {\it not}
the choice of $\varphi$ that we will be particularly interested in.
Instead, we want a $\varphi$ which will have some transparent
connection with the complex structure of ${\bf R}^6$.  If we choose
complex coordinates $z^j = x^j + i y^j$ for $j=1,2,3$, then a D6-brane
in the $x^1$-$x^2$-$x^3$ plane (or any $SU(3)$ image of it) is
calibrated by $\Re\Omega$.  The standard Kahler form on ${\bf R}^6 =
{\bf C}^3$ is
  \eqn{StandardKahler}{
   J = {i \over 2} \left( dz_1 \wedge d\bar{z}_1 + 
    dz_2 \wedge d\bar{z}_2 + dz_3 \wedge d\bar{z}_3 \right) \,,
  }
 and one can readily verify that
  \eqn{TwoPhi}{
    \varphi_0 = \Re\Omega - J \wedge dx^{11}
  }
 is a $G_2$-structure on ${\bf R}^6 \times S^1$ (whose holonomy is
certainly a subgroup in $G_2$).  Actually, much more is true: the
$\varphi_0$ in \TwoPhi\ can be constructed in the same way for any
$CY_3 \times S^1$, and it represents an inclusion of $SU(3)$ in $G_2$.
In fact the Joyce construction we explained in section~\ref{GTwo} is
believed to generalize to ${\bf Z}_2$ orbifolds of $CY_3 \times S^1$,
acting with two fixed points on the $S^1$ and as an anti-holomorphic
involution on the $CY_3$.

The obvious vielbein for $X = {\bf R}^3 \times \hbox{(Taub-NUT)}$ is
  \eqn{ObviousViel}{\eqalign{
   & e^1 = dx^1 \quad e^2 = dx^2 \quad e^3 = dx^3  \cr
   & e^4 = \sqrt{1+H} dy^1 \quad e^5 = \sqrt{1+H} dy^2
      \quad e^6 = \sqrt{1+H} dy^3  \cr
   & e^7 = {1 \over \sqrt{1+H}} \left( dx^{11} + V \right) \,,
  }}
 where 
  \eqn{HDef}{
   dV = *_{y} dH \qquad\hbox{and}\quad H = {R \over 2|\vec{y}|} \,.
  }
 Here $*_{y}$ represents the Hodge dual in the $y^j$ directions only,
and $x^{11} \sim x^{11} + 2\pi R$.  One can now modify $\varphi_0$
slightly to give a $G_2$-structure on $X$:
  \eqn{ThreePhi}{\eqalign{
   \varphi &= dx^1 \wedge dx^2 \wedge dx^3 - 
     (1+H) \left( dx^1 \wedge dy^2 \wedge dy^3 + 
      dy^1 \wedge dx^2 \wedge dy^3 + 
      dy^1 \wedge dy^2 \wedge dx^3 \right) \cr &\qquad{} -
     J \wedge (dx^{11} + V)  \cr
    &= \varphi_0 - H \left( dx^1 \wedge dy^2 \wedge dy^3 + 
      dy^1 \wedge dx^2 \wedge dy^3 + 
      dy^1 \wedge dy^2 \wedge dx^3 \right) - J \wedge V \,.
   }}
 Now, $\varphi \to \varphi_0$ as $|\vec{y}| \to \infty$.  The key
point is that $\varphi_0$ is invariant under $SU(3)$ changes of the
coordinates $z^i$: this is so because both $\Omega$ and $J$ are
$SU(3)$ singlets.  So the $\varphi$ we would construct locally at any
point along the lift of the D6-brane world-volume asymptotes to the
{\it same} $\varphi_0$.  This is the desired verification that the
holonomy of the entire approximation to the seven-dimensional manifold
is (nearly) in the same $G_2$.  I have included the ``(nearly)''
because of the errors in the affine approximation to the D6-brane
world-volume.  This error can be uniformly controlled if there are no
coincident or intersecting D6-branes.  A way to think about it is that
we make $R$ much smaller than the closest approach of one part of the
world-volume to another.

A remarkable fact is that the deformation of the $G_2$-structure,
$\varphi-\varphi_0$, in \ThreePhi, is {\it linear} in $H$.  This is
true despite the fact that the vielbein and the metric are complicated
non-linear functions of $H$.  It is tempting to conjecture that an
appropriate linear modification of $\varphi_0$, along the lines of
\ThreePhi, will be an {\it exact} $G_2$ structure on the whole
seven-manifold, even in cases where D6-branes intersect (of course, in
such a case one must exclude the singularity right at the
intersection).  However, we have been unable to verify
this.\footnote{We thank I.~Mitra, O.~Evnin, and A.~Brandhuber for
discussions on this point.}  Knowing the covariantly constant
three-form suffices to determine the metric, via the explicit formula 
  \eqn{MetricThree}{\eqalign{
   &g_{\mu\nu} = (\det s_{\mu\nu})^{-1/9} s_{\mu\nu} \qquad\hbox{where} \cr
   &s_{\mu\nu} = {1 \over 144} \varphi_{\mu\lambda_1\lambda_2} 
     \varphi_{\nu\lambda_3\lambda_4}
     \varphi_{\lambda_5\lambda_6\lambda_7} 
   \epsilon^{\lambda_1\lambda_2\lambda_3\lambda_4\lambda_5\lambda_6\lambda_7}
   \,,
  }}
 where $\epsilon^{1234567} = \pm 1$, and the sign is chosen to make
$s_{\mu\nu}$ positive definite.  Convincing oneself of the truth of
this formula (which appeared quite early, see for instance
\cite{BryantSal}) is pretty straightforward: $s_{\mu\nu}$ is
symmetric, but scales the wrong way under rigid rescalings of the
manifold to be a metric.  The determinant factor in the definition of
$g_{\mu\nu}$ corrects this problem.  The strategy of finding $\varphi$
first and then deducing $g_{\mu\nu}$ has been of use in a recent
investigation in the string theory literature \cite{brandhuber}.

\section*{Acknowledgments}

I thank A.~Brandhuber, J.~Gomis, S.~Gukov, and E.~Witten for extensive
discussions about $G_2$ holonomy, and J.~Schwarz for a careful reading
of the manuscript.  I would like to express my special appreciation to
K.~T.~Mahanthappa for his many years of dedication to TASI, and to
Kathy Oliver for her help in making the school function.  This work
was supported in part by the DOE under grant DE-FG03-92ER40701.

\bibliography{tasi}
\bibliographystyle{ssg}      

\end{document}